\RequirePackage{fix-cm}
\documentclass{svjour3}                     % onecolumn (standard format)
\smartqed  % flush right qed marks, e.g. at end of proof
\usepackage[bookmarks=false]{hyperref}
\usepackage{gensymb,hyperref,graphicx,array,amsmath,caption,epstopdf,amsfonts,algorithm,algorithmic}
\usepackage{subfigure,psfrag,epsfig,amssymb,array,flushend,xcolor,color}
\begin{document}
%\tableofcontents
\title{Hybrid  blind robust image watermarking technique based on DFT-DCT and Arnold transform %\thanks{Grants or other notes
}


\author{Mohamed Hamidi \and Mohamed El Haziti  \and Hocine Cherifi   \and Mohammed El Hassouni
}


\institute{M. Hamidi  \and M. EL HAZITI \and M. EL HASSOUNI\at
LRIT-CNRST URAC 29, Rabat IT Center, Faculty of Sciences,\\ Mohammed V University in Rabat, Morocco \\             % Tel.: +123-45-678910\\
                         %  \\
           % \emph{Present address:} of F. Author  
            %  if needed
		  \and M. Hamidi \\
		  \email{hamidi.medinfo@gmail.com} \\
		 M. El Hassouni \\
	LRIT-CNRST URAC 29, Rabat IT Center, FLSH, \\ Mohammed V University in Rabat, Morocco\\
	\email{mohamed.elhassouni@gmail.com} \\	  
			 M. El Haziti \at
              LRIT - CNRST URAC 29, Rabat IT Center, EST,\\ Mohammed V University in Rabat, Morocco\\
\email{elhazitim@gmail.com}
			\and
           H. Cherifi \at
      Laboratoire Electronique, Informatique et Image (Le2i) UMR 6306 CNRS,\\ University of Burgundy, Dijon, France \\
\email{hocine.cherifi@u-bourgogne.fr}
}

\date{Received: date / Accepted: date}

\maketitle

\begin{abstract}

In this paper, a robust blind image watermarking method is proposed for copyright protection of digital images. This hybrid method relies on combining two well-known transforms that are the discrete Fourier transform (DFT) and the discrete cosine transform (DCT). The motivation behind this combination is to enhance the imperceptibility and the robustness. The imperceptibility requirement is achieved by using magnitudes of DFT coefficients while the robustness improvement is ensured by applying DCT to the DFT coefficients magnitude. The watermark is embedded by modifying the coefficients of the middle band of the DCT using a secret key. The security of the proposed method is enhanced by applying Arnold transform (AT) to the watermark before embedding. Experiments were conducted on natural and textured images.  Results show that, compared with state-of-the-art methods, the proposed method is robust to a wide range of attacks while preserving high imperceptibility.

%Moreover, the proposal shows an improvement over the alternative techniques to a wide range of common attacks while preserving high imperceptibility.
%\textcolor{red}{
%The main purpose of the proposed method is to take the advantages of jointing two well known transforms in order to enhance imperceptibility and robustness. The imperceptibility requirement is achieved by modifying the discrete Fourier transform (DFT) magnitude instead of the DFT phase. For more robustness the discrete cosine transform (DCT) is applied to the DFT magnitude.  The underlying idea is to carry out a discrete cosine transform (DCT) to the magnitude resulting from a discrete Fourier transform (DFT) applied to the host image. Then, the watermark is embedded by modifying the middle band coefficients of the DCT using a secret key. The security of the proposed system is enhanced by using Arnold Transform before the embedding stage. Experimental results, compared with state-of-the-art methods, show the effectiveness of the proposed method in terms of  imperceptibility and robustness. Moreover, the proposal shows an improvement over the alternative techniques to a wide range of common attacks while preserving high imperceptibility.}
%%, e.g., JPEG compression, histogram equalization, low-pass Gaussian filtering, Gaussian noise, salt \& pepper noise, Gaussian smoothing, cropping and the combination of some attacks
\keywords{  image watermarking \and Copyright protection \and hybrid method \and Discrete Fourier Transform (DFT) \and Discrete Cosine Transform (DCT) \and Arnold transform}
\end{abstract}

\section{Introduction}
\label{intro}
With the rapid growth of Internet and new technologies for multimedia services and the proliferation of digital devices, multimedia data can be modified, duplicated and distributed very easily. Therefore, preventing unauthorized use of these contents has become more and more important.
% Hence, the strong demand for reliable and secure protection techniques for multimedia data. 
 To overcome this issue, digital image watermarking, especially robust image watermarking, is an efficient solution. The underlying concept  of image watermarking is to embed a watermark within the cover image to protect it from illegal usage. The watermark must be imperceptible, so that it  should not degrade the quality of the host image and it should be difficult or even impossible to counterfeit or remove it. 

In general, the process of image watermarking must satisfy four requirements which are imperceptibility, robustness, capacity and security \cite{RefB1}. A good watermarking scheme must provide the best tradeoff between these four properties according to the requirements of the aimed application. The first important requirement of an image watermarking system is imperceptibility. It refers to perceptual similarity between the original image and the watermarked image. Indeed, an efficient watermarking scheme should produce no artifacts or quality loss in the images. If the watermarking scheme fails to achieve this requirement, it will not be suitable for practical applications. The second property is the robustness. It represents the ability of detecting the watermark even if the watermarked image has incurred changes in its distribution process. Consequently, the watermark needs to be robust against common signal processing operations such as filtering, noise addition, lossy compression, cropping, etc. The third requirement is the capacity which refers to the maximum number of bits that can be embedded in a given host data.
The fourth requirement is the security of watermark. It refers to its ability to resist hostile attacks, so that unauthorized users cannot remove the watermark. In order to achieve a minimum level of security,  a secret key is required in watermarking systems.

% Note that the requirements on these properties are often contradictory. For instance, by increasing the capacity, the imperceptibility will increase and the robustness will decrease and vice versa.
 
 Digital image watermarking can be used in a wide variety of applications. In Copyright protection, the goal is to secure digital images in unsecured networks like Internet. The ownership can be proven in the case of dispute, by extracting the owner's copyright information embedded invisibly into the host image. Authentication is also an interesting application of image watermarking which aims to detect if any modification has been applied to the host image and then localizes exactly the tampered region. Another application is tamper detection. The presence of tampering is achieved by embedding a fragile watermark. If the watermark is degraded or destroyed, it indicates that the image cannot be trusted. This process is used in applications involving sensitive data such as medical imagery, satellite imagery, etc. 
% Whereas, broadcast monitoring refers to the process of verifying where and when an advertisement was broadcasted. The verification is done by recognizing watermarks embedded in the contents that were supposed to be broadcasted. 
Machine learning techniques can be widely used in several applications such as recognition \cite{RefReviewer2}\cite{RefReviewer3}. In digital  image watermarking, machine learning approaches and artificial intelligence mechanisms such as neuro computing, fuzzy techniques
as well as evolutionary algorithms are used in the watermarking field \cite{MachLearning}.

 There are different classification  criteria for watermarking systems.
 Based on the resistance to attacks, watermarking algorithms are divided into three main categories; fragile \cite{RefJ1}, semi-fragile \cite{RefJ2} and robust watermarking \cite{RefJ3}.
 Fragile watermarking schemes have been proposed especially for image authentication and integrity verification. 
They are used to detect any unauthorized modification at all. Semi-fragile techniques are implemented for detecting any unauthorized modification, while allowing  at the same time some image-processing operations.  
Robust watermarking algorithms are designed to survive arbitrary, malicious attacks such as image scaling, cropping, and lossy compression. They are usually used for copyright protection with the aim of declaring rightful ownership.  
%
% These schemes embed an imperceptible watermark in host image which is difficult to counterfeit without authorization, and allows for authorized detection of tampering, preferably with localization information.
%  
  The existing algorithms can be also classified into spatial and transform domains. Spatial domain techniques \cite{SpatialMethod2017} embed the watermark by directly modifying the image pixels, whereas in frequency domain techniques \cite{RefJ5} a transformation is first performed and then   the watermark is embedded into discrete cosine transform (DCT) \cite{RefJ19}\cite{DCTinterblock2014} \cite{SinghMedicine2016}\cite{SinghFutureGeneR2016}, discrete wavelet transform (DWT) \cite{RefJ6}\cite{SinghWavelet2015} \cite{SinghTelemedicine2015}\cite{SinghAK2015} or  discrete Fourier transform (DFT) coefficients \cite{RefJ7}\cite{RefJ8}.  
  For applications, such as authentication, tamper detection, copyright protection,  it is desirable to be able to extract the watermark without the original image. This requirement, introduces a very challenging issue especially if robustness is also needed. We distinguish between non-blind \cite{RefNonBlindMethod}, semi-blind \cite{RDWTMTAP2017} and blind \cite{RefJ11} \cite{Roy2016} watermarking systems depending on whether or not the host image is needed during watermark extraction. In non-blind techniques, the original image is needed; Semi-blind methods require the watermark and some side information; Blind approaches neither need the original image nor the watermark. 
  \par
% Finally, hybrid techniques take into account both the frequency and spatial characteristics of the watermark, e.g., by modifying the block DCT coefficients of the image [11] or by operating in a space-scale domain like the wavelet domain [13][14].\\
In this paper, we propose a blind robust image watermarking technique for copyright protection. 
The watermark is embedded in middle band coefficients of DCT of the magnitude after carrying out the DFT of the original image.
The choice of using DFT magnitude has been driven by the gain in terms of imperceptibility. However, it has been found that the robustness of the proposed scheme is weak when the DFT magnitude is used only. To overcome this problem, we choose to apply the DCT to the DFT magnitude thanks to its advantages especially its robustness against signal processing attacks. 
% Thanks to the advantages that the DCT offer especially its robustness against common signal processing attacks, we choose to apply the DCT to the DFT magnitude.
 In addition, to enhance the security of the proposed method, the Arnold transform is used to encrypt the watermark. The gain obtained after jointing these two transforms in terms of imperceptibility and robustness is clearly illustrated in experimental results. 

To evaluate the proposed scheme, we compare its DFT counterpart in terms of imperceptibility and robustness. Furthermore, comparative experiments are performed with alternative methods presented in 
%\cite{lien2006watermarking},
 \cite{RefJ14}, \cite{RefJ15}, \cite{RefJ16},\cite{RDWTMTAP2017}, \cite{SinghAK2015} and \cite{SinghAK2016}.
%The proposed method is inspired by the experiment proposed by Oppenheim \textit{ et al.}  \cite{RefJ13} that shows the importance of the phase compared to the magnitude.\\
%\textcolor{cyan}{
%The idea behind embedding the watermark in magnitude of the DFT is that it has less effect on the host image, whereas a small change of the DFT phase can cause large changes to the original image \cite{RefJ14}. 
%With the aim of increasing the imperceptibility and the robustness to common signal processing operations, we apply the DCT to the DFT magnitude.

% More specifically, the watermark is embedded in middle band coefficients of DCT of the magnitude after carrying out the DFT of the original image.

% As a result of exploiting the advantage of DFT magnitude and  DCT transform, we obtain a method which ensures high imperceptibility and which is robust to a wide range of attacks. 
 
% Moreover, we use the Arnold transform to encrypt the watermark in order to increase the security of the proposed scheme.%}
 \par
 This paper is organized as follows. Section \ref{Related work} discusses the related works. Section \ref{Preliminaries} gives a description of used terminologies followed by the proposed watermarking scheme illustrated in Section \ref{Proposed scheme}. Section \ref{Experimental setup} sketches the experimental setup. Section \ref{Experimental Results} reports the experimental results. Finally, Section \ref{Conclusion} concludes the paper.
\section{Related work}
\label{Related work}
%Due to the robustness of transform domain approaches
% %\cite{RefJ18}
%, researchers turned their attention to frequency domain methods rather than the spatial domain techniques for robust watermarking, especially, with the aim of Copyright protection \cite{RefJ19} .\par
In the literature, several watermarking methods have been proposed in the transform domain. 
One of the most popular watermarking scheme is introduced in \cite{RefJ5}, by Cox \textit{et al.} where a pseudo-random Gaussian sequence is embedded into the largest $1000$ AC coefficients in the DCT domain.  
%\textcolor{blue}{
%One of the most popular watermarking schemes is introduced in \cite{RefJ5}, by Cox \textit{et al.} . It is based on Spread Spectrum Communication. A pseudo-random Gaussian sequence is embedded into the largest $1000$ AC coefficients in the DCT domain. This method is robust to common image processing and geometric distortions. 
%}
%Enlever la méthode de Barni
%\textcolor{blue}{
%Barni \textit{et al.} \cite{RefJ21} proposed a perceptually tuned scheme using additive modulation. A pseudo-random sequence of real numbers is embedded in a selected set of DCT coefficients. In order to  ensure the watermark invisibility it is tailored to the image by exploiting the masking characteristics of the human visual system (HVS). The  watermark detection is accomplished by computing a correlation measure.
%}
%\textcolor{blue}{
%In \cite{RefJ19}, Hsu et \textit{al.} proposed a DCT based watermarking scheme. The watermak is embedded in the middle frequency part of the image. Similarly to our approach, the authors carry out a pretreatment of the watermark by applying a pseudo-random permutation to enhance the security. This method is robust against image processing operations and JPEG lossy compression. However, the geometric attacks are still challenging to this work.  
Das \textit{et al.} in \cite{DCTinterblock2014} proposed a DCT watermarking method based on correlation between DCT coefficients in the same position of adjacent blocks.

%Pun \textit{et al.}  \cite{RefJ20} has proposed a  DFT-based watermarking system for images.  The original image is decomposed into Fourier domain, then the watermark is embedded in the Fourier coefficients with highest magnitudes.
 Poljicak \textit{et al.} \cite{Ante2011} the watermark is inserted in the magnitude of the Fourier transform  taking the advantage of minimizing the impact of the watermark implementation on the overall quality of an image.
In \cite{RefJ14}, Wang \textit{et al.} proposed a wavelet-tree-based blind watermarking scheme for copyright protection. The watermark bits are embedded by quantizing super trees.  
%As the watermark, is spread throughout large spatial regions, robustness against several kind of attacks increases. 
%After that, Lien and Lin \cite{lien2006watermarking} also improved Wang and Lin???s  method \cite{wang2004wavelet} by using the wavelet tree. 
In \cite{RefJ15}, a blind watermarking method based on quantization of distance between wavelet coefficients is proposed. 
%A set of coefficients are quantized according to the watermark bits.
%First, the wavelet coefficients are divided into blocks and the first, the second and the third maximum coefficients in each blocks are obtained. Next, the first and the second maximum coefficients are quantized according to the watermark bits. 
%Finally, the extracting process is performed using the block-based watermarking. This method is robust against 
%common image processing attacks, especially low pass filtering and JPEG compression.
In \cite{RefJ16},Lin \textit{et al.} proposed a wavelet-tree-based watermarking method using distance vector of binary cluster for copyright protection. The watermark is embedded into insignificant coefficients of a wavelet tree.
%  First, the authors use the distance vector to denote binary watermark bits and the wavelet trees are classified into two clusters.  Second, the two smallest wavelet coefficients in a wavelet tree are used  with the aim of reducing distortion of a watermarked image. 
% 
% Third, the distance vector is quantized to decrease the image distortion. Next, the trees are classified into two clusters in such a way that they exhibit a large statistical difference based on the distance vector. 
%
%Finally, a comparison between the statistical difference and the distance vector of a wavelet tree is carried out to extract the correct watermark bit. The method is robust against filtering, jpeg compression, etc.  

Note that all these solutions are based on a single transform domain and that they all try to insert the watermark in selected values in order to increase robustness and imperceptibility. 
The main motivation of the majority of  existing watermarking schemes is to improve the robustness  against a wide range of attacks while preserving a good visual quality of images. Therefore, the need to develop hybrid methods that combine two or more transforms to use the characteristics of these transforms and achieve the required aims has increased considerably \cite{MedicalImagesAmit2017}. \\
%\textcolor{magenta}{
In \cite{RefJ24} a digital image watermarking scheme based on DCT and SVD is proposed. This approach used differential evolution (DE) to select adaptively the strength of the watermark and Arnold transform (AT) in order to enhance security. First, the host image is divided into $8 \times 8$ square blocks and then the DCT is applied  on each block. Afterwards, the DC components of DCT coefficients are collected with the aim of constructing the low resolution approximation matrix. Finally, the watermark image is scrambled using the Arnold transform (AT) and embedded into the diagonal matrix S using a scaling factor obtained with a differential evolution algorithm.
In \cite{RefJ25}, a blind watermarking algorithm based on Discrete Wavelet Transform (DWT) and Discrete Cosine Transform (DCT) is proposed. The watermark is scrambled by Arnold transform and embedded in a spread spectrum pattern using pseudo random in the mid frequency coefficients of the corresponding DCT blocks of  DWT LL sub-band. Experimental  results show that combining the two transforms gives better results than using DCT only.
%}
%\textcolor{magenta}{
In \cite{RefJ26}, a blind robust image watermarking method based on DWT-SVD and DCT using Arnold Cat Map encryption for copyright protection is proposed.  The DCT coefficients of the watermark image  are embedded into the middle singular value of each block having 
size $4 \times 4$ of the host image's one level Discrete Wavelet Transform (DWT) sub-bands.
This scheme  is secure, imperceptible and robust against common signal processing operations.

In \cite{Amit2014}, Amit Kumar Singh \textit{et al.}  proposed a hybrid method based on DWT-DCT-SVD. The original image is decomposed into first level DWTs. Next, the DCT and SVD are applied to the low frequency band (LL). Afterwards, the watermark image is transformed also using the DCT and SVD. Then, the S component of watermark is inserted in the S component of the host image. The method is robust against signal processing attacks. 

In \cite{Roy2016}, Soumitra Roy \textit{et al.} proposed a RDWT-DCT based blind image watermarking scheme using Arnold scrambling. First, the original image is decomposed into non overlapping blocks and the RDWT (Redundant Discrete Wavelet Transform) is carried out to each block. Second, the watermark is encrypted using Arnold chaotic map to increase the security.
Then, the DCT is applied to each LH subband of the non-overlapping host image block. Finally, the watermark is embedded by modifying middle significant AC coefficients using repetition code. Soumitra's method is shown to be robust against geometric attacks, jpeg compression among others.

In \cite{RDWTMTAP2017}, a semi-blind gray scale image watermarking technique in redundant wavelet domain using the combination of non sub-sampled contourlet transform(NSCT), redundant discrete wavelet transform(RDWT) and SVD is proposed. Singh's \textit{et al.} used Arnold transform encryption to enhance the security of the watermarking system.  This method is shown to be robust to geometrical and signal processing attacks.
In \cite{SinghAK2015}, a secure DWT-DCT-SVD based image watermarking technique is proposed. First, the host image is decomposed up to second level of DWT. Second, the DCT and SVD have been applied to the low frequency ll of  the original image. The watermark medical image is also transformed by DCT and SVD. The watermark embedding is performed by inserting the singular value of watermark image in the singular value of the original image. In \cite{SinghAK2016}, an hybrid image watermarking technique based on Nonsubsampled contourlet transform(NSCT), Multiresolution Singular value decomposition (MSVD), discrete cosine transform (DCT) and Arnold transform is proposed. In this method, three image watermarks have been added into the cover image exploiting the advantages of combining  these transforms to enhance robustness, capacity and imperceptibility requirements.

Recently, an interesting survey of image watermarking techniques and their application have been proposed \cite{SinghAK2018}. Kumar \textit{et al.} have discussed recent state-of-art watermarking techniques issues and potential solutions. The work can be useful for secure e-governance applications.

The major limitation of the existing image watermarking schemes for copyright protection is the difficulty to ensure a good tradeoff between imperceptibility and robustness. To take full advantage of image transforms, we design a novel combination of DFT and DCT for blind robust image watermarking. The reason behind this choice is due to the fact that the DFT magnitude shows ability to ensure high imperceptibility while the DCT can improve the robustness of the proposed technique to common signal processing attacks. Furthermore, Arnold transform is used to enhance the security of the proposed watermarking system. 
%Inspired by these works, we propose to exploit the effectiveness of the two transform approaches using a watermark encryption to enhance the security of the proposed system.
%\label{sec:1}
\section{Preliminaries}
\label{Preliminaries}
\subsection{Discrete Fourier transform (DFT)}
Discrete Fourier transform of an image leads to magnitude and phase representation. This transformation has several characteristics. An important property of the DFT is its translation invariance. In fact, spatial shifts doesn't affect the magnitude but affect the phase component \cite{RefJ27}. DFT is also robust to cropping. In fact, when the watermark is embedded in the magnitude, even if the spectrum is blurred, the synchronization is not needed.
The discrete Fourier transform of an image $f(x,y)$ of size $M \times N$ and the inverse DFT (IDFT) are defined respectively as follows : 
\begin{equation}
\begin{split}
F(u,v)= \frac{1}{MN} \sum_{x=0}^{M-1}\sum_{y=0}^{N-1}f(x,y)e^{-2\pi j(\frac{ux}{M}+\frac{vy}{N}) }  \\ = R(u,v)+jI(u,v)
\end{split}
\end{equation}
\begin{equation}
f(x,y)= \sum_{u=0}^{M-1}\sum_{v=0}^{N-1}F(u,v)e^{2\pi j(\frac{ux}{M}+\frac{vy}{N}) }
\end{equation}
Where $R(u,v)$ and $I(u,v)$ are the real and the imaginary parts of the Fourier transform, respectively. 
Equation shows the polar representation of the Fourier transform :
\begin{equation}
F(u,v)= \left | F(u,v) \right |e^{j\phi (u,v)}
\end{equation}
Where $\left | F(u,v) \right |$ and $\phi(u,v)$ are respectively the Fourier magnitude and the Fourier phase. They are represented as follows : 
\begin{equation}
\label{eq:Magnitudeformula}
M(u,v)= \left | F(u,v) \right | = [R^{2}(x,y)+I^2(x,y)]^{1/2}
\end{equation}
\begin{equation}
\label{eq:Phaseformula}
\phi (u,v)= tan^{-1}\left [ \frac{I(u,v)}{R(u,v)} \right ]
\end{equation}
Where $R(u,v)$ and $I(u,v)$ are respectively the real and imaginary parts of $F(u,v)$.
\subsection{Discrete cosine transform (DCT)}

The discrete cosine transform (DCT) is one of the famous transformation technique that transforms an image from the spatial domain to the frequency domain \cite{RefJ28}. It has been widely applied in image processing exploiting both the decorrelation and the energy compaction properties. Generally, DCT watermarking approaches used square matrix of $8 \times 8$ as block size. The mathematical expressions of the 2D-DCT and inverse 2D-DCT are respectively : 
\begin{equation}
\begin{split}
C(u,v)=\frac{2}{\sqrt{mn}} \alpha(u)\alpha (v) \sum_{x=0}^{M-1}\sum_{y=0}^{N-1}f(x,y)\times \\ cos\frac{(2x+1)u\pi }{2m} \times cos\frac{(2y+1)v\pi }{2n}
\end{split}
\end{equation}
\begin{equation}
\begin{split}
f(x,y)=\frac{2}{\sqrt{mn}}  \sum_{u=0}^{M-1}\sum_{v=0}^{N-1}\alpha(u)\alpha (v) C(u,v)\times \\ cos\frac{(2x+1)u\pi }{2m} \times cos\frac{(2y+1)v\pi }{2n}
\end{split}
\end{equation}
Where $f(x,y)$ and $C(u,v)$ are respectively  the pixel values in the spatial domain and the DCT coefficients. $m,n$ represent the block size.
\begin{equation}
\alpha(u)\alpha (v)  = \left\{\begin{matrix}
\frac{1}{\sqrt{2}} & if (u,v)=0 \\ 
 1 & else
\end{matrix}\right.
\end{equation}

\subsection{Arnold Transform}
%Digital image scrambling is a preprocessing during hiding a watermark into a cover image.
 The idea behind using scrambling algorithm is to enhance the security of the watermark in order to avoid unauthorized person to counterfeit or remove it. Therefore, it guarantees more safety and reliability for the image in the transmission process. Arnold scrambling is widely used  in digital image watermarking due to its simplicity and periodicity \cite{RefJ29}\cite{ArnoldNonSquare}.  According to this periodicity, after several cycles, the host image can be easily restored. The obtained watermark image after the scrambling process is chaotic, so without the scrambling algorithm and the key, the attacker cannot decrypt it even if it has been extracted from the watermarked image. Moreover, the spatial  relationships between the pixels have been destroyed which ensures more security. We note that the period of Arnold transform should less than $\frac{N^2}{2} $ with N the image size \cite{RefJ29}.

 The result of applying the Arnold transform is randomly organizing the pixels of the image. The principle idea is that if iterated enough times, the host image reappears. Note that the parameters of Arnold transform serve as the additional key for enhancing the security.

Arnold transform, also called cat map transform, is defined as :
\begin{equation}
\label{eq:Arnoldtransform}
\begin{bmatrix}
a'\\ 
b'
\end{bmatrix} = \begin{bmatrix}
1 & 1 \\ 
1 & 2
\end{bmatrix} \begin{bmatrix}
a\\ 
b
\end{bmatrix} mod (n)
\end{equation}
Where $(a,b)$ and $(a',b')$ are the pixel coordinates of the original watermark and the encrypted watermark, respectively. Let $A$ the left matrix in the right part of equation (\ref{eq:Arnoldtransform}), $I(a, b)$ and $I(a', b')^{(n)}$ represent pixels in the original watermark and the encrypted one obtained by performing Arnold transform $n$ times, respectively.
Thus, using $n$ times the Arnold transformation, the watermark encryption can be written as :
\begin{equation}
\label{eq:Watermarkencryption}
I(a', b') ^{(p)} = A I(a,b)^{(p - 1)} mod (n) 
\end{equation}
Where $p = 1, 2, \dots , n$, and $I(a',b')^{(0)} = I(a, b)$.
In fact, to obtain $I(a,b)^{(p - 1)}$ we multiply the inverse matrix of $A$ at each side of equation (\ref{eq:Watermarkencryption}). That is to say, by iteratively calculating the formula (\ref{eq:Watermarkdecryption}) $n$ times, the encrypted watermark can be decrypted easily.
 \begin{equation}
\label{eq:Watermarkdecryption}
J(a, b)^{(p)} = A^{-1}J(a', b')^{(p - 1)}mod (n) 
\end{equation}
Where $J(a', b')^{(0)}$ is a pixel representation of the encrypted watermark  and $J(a,b)^{(p)}$ is the decrypted pixel after $p$ iterations. 
\begin{figure}[!h]
    \centering
    \subfigure[]{\label{sub21} \includegraphics[width=0.30\textwidth]{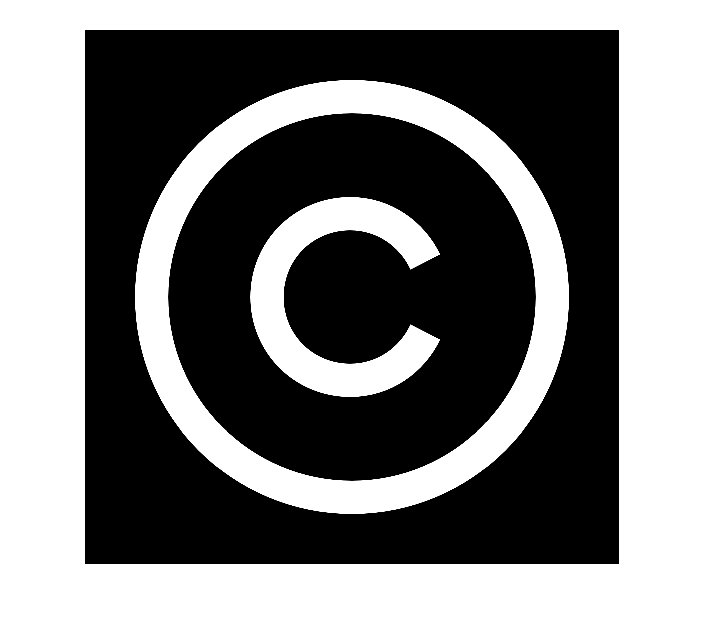}}
    \subfigure[]{\label{sub22} \includegraphics[width=0.30\textwidth]{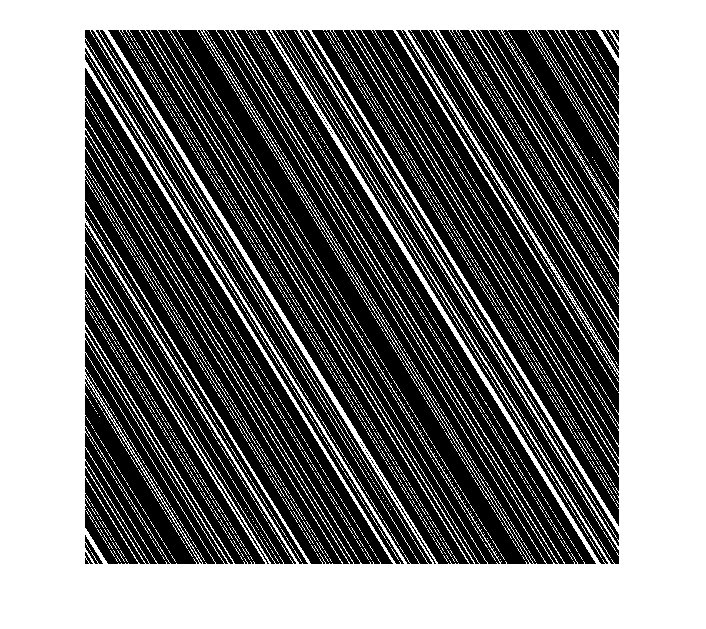}}
    \caption{(a) Original watermark, (b) Encryped watermark}
    \label{fig:Arnoldtransf.}
    \end{figure}
    Figure \ref{fig:Arnoldtransf.} shows the watermark image encryption using Arnold transforms, where (a) is the original watermark sized $64 \times 64$ , and (b) is the encrypted one with $n = 50$.
\section{Proposed scheme}
\label{Proposed scheme}

In this work, we propose a blind robust image watermarking method for copyright protection. 
The watermark is inserted in  the DCT middle band of the  DFT magnitude. The reason behind the choice of DFT magnitude has been driven by the gain in terms of watermark imperceptibility. Nevertheless, the scheme shows robustness  weakness when the DFT magnitude is used only. Since the DCT is very robust against signal processing attacks, we believe it is an excellent solution to consider in our scheme. For this reason, the DCT is applied to the DFT magnitude to enhance the watermark robustness. Furthermore, the watermark is encrypted with the Arnold transform to increase the security of the proposed method.

%\textcolor{blue}{
% It is well recognized from the experiment proposed by Oppenheim \textit{et al.} that the information carried by the phase of the image appears to be much more significant than the magnitude \cite{RefJ12}. Consequently, the most important features of an image are preserved if the phase is kept untouched. 
% For this reason that we propose to insert the watermark in the magnitude of the DFT instead of the phase. However, it has been found that the robustness of the proposed scheme is weak when the DFT magnitude is used only. 
% In order to increase the robustness, we apply the DCT to the DFT magnitude thanks to its robustness against several attacks especially signal processing operations. Thus, the watermark is embedded in the  mid-band DCT  coefficients of the DFT magnitude using a secret key (key 1). Moreover, a second key (key 2), which refers to the parameters of Arnold transform, is used as additional key with the aim of enhancing security.  
% By jointing two transforms, namely the DFT and DCT, we design an image watermarking method which is robust to a wide range of attacks while preserving high imperceptibility.} 
%  They proved that the information carried by the phase of the image appears to be much more significant than that carried by the magnitude. 
% Thanks to the fact that the modification of the magnitude does not influence the image compared to the influence of the phase, we exploit this fact to  insert the watermark without altering the image.
\subsection{Watermark embedding}
\begin{figure}[!t]
%\centering
\captionsetup{justification=centering}
\begin{center}
\includegraphics[width=0.75\textwidth]{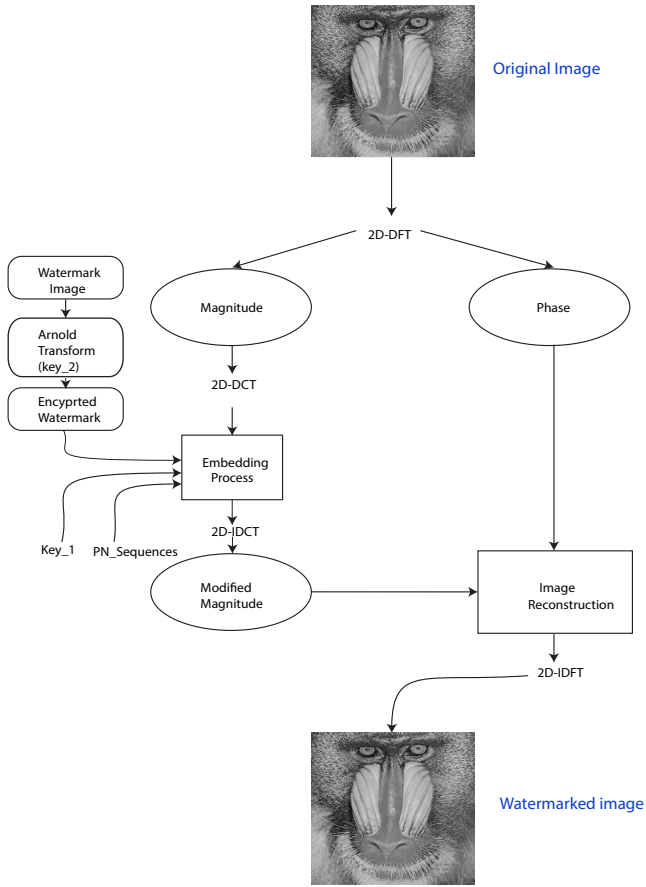}
\end{center}
 \caption{The embedding process of the proposed scheme}
\label{fig:EmbeddingScheme}
\end{figure}

%\begin{figure}[!h]
%%\centering
%\captionsetup{justification=centering}
%\begin{center}
%\includegraphics[scale=0.55]{figs/dct_frequencies.eps}
%\end{center}
%\caption{The middle band DCT coefficients where the watermark is embedded.}
%\label{fig:middlebandfrequencies.}
%\end{figure}
%The pseudo-random sequence can take only two values  ??????$\begin{Bmatrix} -1,1 \end{Bmatrix}$.
% The embedding process consists of embedding PN-Sequences depending on the bit of the watermark. The middle band coefficients (see the colored region in Fig. \ref{fig:middlebandfrequencies.}) of the DCT transform of the DFT magnitude are commonly used for watermark embedding to avoid modifiying the important visual parts of image.\par
%   \\ The proposed watermark embedding scheme is detailed in \textbf{ Algorithm 1}.
%Ajouter aussi dabord pourquoi choisir Arnold anisi que ses parametres qui sont pris comme une cl?? secrete (key 2) et dire qu'on a apres cette AT une marque crypt??e ce qui renforce la s??curit??.
The proposed embedding mechanism is illustrated in Fig. \ref{fig:EmbeddingScheme}. 
First of all, a prepossessing is applied to the watermark. It consists on Arnold transformation which is used to encrypt the watermark as a security enhancement. The idea behind this transformation is to make difficult even impossible for an unauthorized person, without knowing the scrambling algorithm and parameters that represent an additional secret key (key 2),  to detect the real watermark even if it's correctly extracted. 
Initially, the DFT is applied to the original image then the the magnitude $M(u,v)$ and the phase $\phi (u,v)$   are calculated using equations (\ref{eq:Magnitudeformula}) and (\ref{eq:Phaseformula}). Afterwards, the magnitude matrix is divided into square blocks of size $8 \times 8$. Then,  the DCT is computed on every block of the magnitude. Next, using the secret key 1, two uncorrelated pseudo-random sequences are generated : one sequence for  "0" bits (PN\_Seq\_0) and another  sequence for the "1" bits (PN\_Seq\_1).  Note that PN sequences must have the same size than the middle band coefficients. \par
%The pseudo-random sequence can take only two values  ??????$\begin{Bmatrix} -1,1 \end{Bmatrix}$.
 The embedding process consists of inserting the PN-Sequences according to the bit value of the watermark using equation \ref{eq:Embeddingformula}.
 $M(u,v)$ is obtained after applying the DCT to the DFT magnitude.  The watermark $W(u,v)$ that consists of two PN sequences, is inserted in the middle band coefficients. The strength of embedding is adjusted by the parameter $k$ which controls the tradeoff  between the robustness and imperceptibility. 
  The middle band coefficients of the DCT transform of the DFT magnitude are  used for watermark embedding to avoid modifying the important visual parts of image. The original DFT magnitude and the modified one are depicted in Fig. \ref{fig:OriginalAndModifiedMagnitude}. 
\begin{equation}
\label{eq:Embeddingformula}
M_w(u,v)=\begin{Bmatrix}
M(u,v)+k*W(u,v) \hspace*{0.5cm} u,v \in F_M \\ 
M(u,v) \hspace*{2.5cm} u,v \notin F_M         
\end{Bmatrix}
\end{equation}
 Where  $M_w(u,v) $ is the watermarked magnitude block, $M(u,v)$  represents the $8 \times 8$ DCT block of DFT  magnitude, $W(u,v)$ represents the watermark which consists of two PN sequences, $F_M$ refers to the middle frequency band  which is modified during the watermark embedding, and $k$  is the watermark strength that controls the tradeoff between imperceptibility and robustness requirements. 
 \par
 
%\begin{figure}[!t]
%\centering
%\captionsetup{justification=centering}
%\begin{center}
%\includegraphics[width=0.75\textwidth]{figs/OriginalMagnitude.eps}
%\end{center}
%\caption{ Original Magnitude of D9}
%\label{fig:OriginalMagnitude}
%\end{figure}
%Explanation why we choose PN SEQUENCES etc
    The watermark consists of two pseudo-random sequences PN\_Seq\_0 and  \\ PN\_Seq\_1 (see Algorithm \ref{alg:embedding-process}). Each sequence is a vector composed by $\left \{ -1, 1 \right \}$ values with a normal distribution having zero mean and unity variance. The motivation behind this choice (normally distributed watermark) is the robustness to the attacks  trying to produce an unwatermarked document by averaging multiple differently watermarked copies of it \cite{RefPN}. In the detection side it is important that the PN sequences are statistically independent. This constraint is granted by the pseudo-random nature of the sequences. In addition, such sequences could be easily regenerated by providing the correct seed (key 1).

  The watermark strength is handled by the gain factor $k$ which controls the tradeoff between robustness and imperceptibility. In fact, an increase of the gain factor increases the watermarking robustness while it decreases the imperceptibility of the watermark. Thus, we choose empirically the value of $k$ so that we have a good tradeoff between robustness and imperceptibility. After, inverse DCT is applied to obtain the modified  magnitude. Finally, the watermarked image is reconstructed with the unchanged phase and the modified magnitude using equation (\ref{eq:Imagereconstruction}). 
\begin{equation}
\label{eq:Imagereconstruction}
\hspace*{-1cm}
I_w(u,v) = M_w(u,v)*e^{(j\phi (u,v))}
\end{equation}  
  Afterwards, the inverse discrete Fourier transform (IDFT) is performed to obtain the watermarked image.\\
Fig. 2 sketches the watermark embedding process which is described in detail in Algorithm \ref{alg:embedding-process}.  
\begin{algorithm}[h]
\caption{Watermark embedding}
\label{alg:embedding-process}
\begin{algorithmic} 
\REQUIRE Original image, Watermark, key 1, key 2, PN\_Seq\_0, PN\_Seq\_1.
\ENSURE Watermarked image.
\STATE 1. Perform Arnold transform to the watermark image using the secret key 2.
\STATE 2. Apply DFT to the original image and calculate the magnitude and the phase.
\STATE 3. Generate two uncorrelated PN sequences for middle frequency band coefficients using the secrete key 1. 
\STATE 4. Divide the magnitude of DFT into $8 \times 8$ blocs and then apply the DCT.
\STATE 5. Insert the two PN sequences bits according  to watermark bits using the equation (\ref{eq:Embeddingformula}) \\
 \textbf{if} Watermark (bit) $= 0$ \hspace*{0.1cm}  \textbf{then}  \hspace*{0.1cm} $W(u,v)=$ PN\_Seq\_0\\ \hspace*{0.2cm} \textbf{else}  \hspace*{0.5cm} $W(u,v)=$ PN\_Seq\_1.
\STATE 6. Perform IDCT on each watermarked magnitude block.
\STATE 7. Reconstruct the watermarked image with the modified magnitude using the equation (\ref{eq:Imagereconstruction}) \cite{RefJ12}. 
\STATE  8. The final watermarked image is obtained  by performing the  IDFT.
\end{algorithmic}
\end{algorithm}

\subsection{Watermark extraction}
\begin{figure}[t]
\centering
\includegraphics[width=0.75\textwidth]{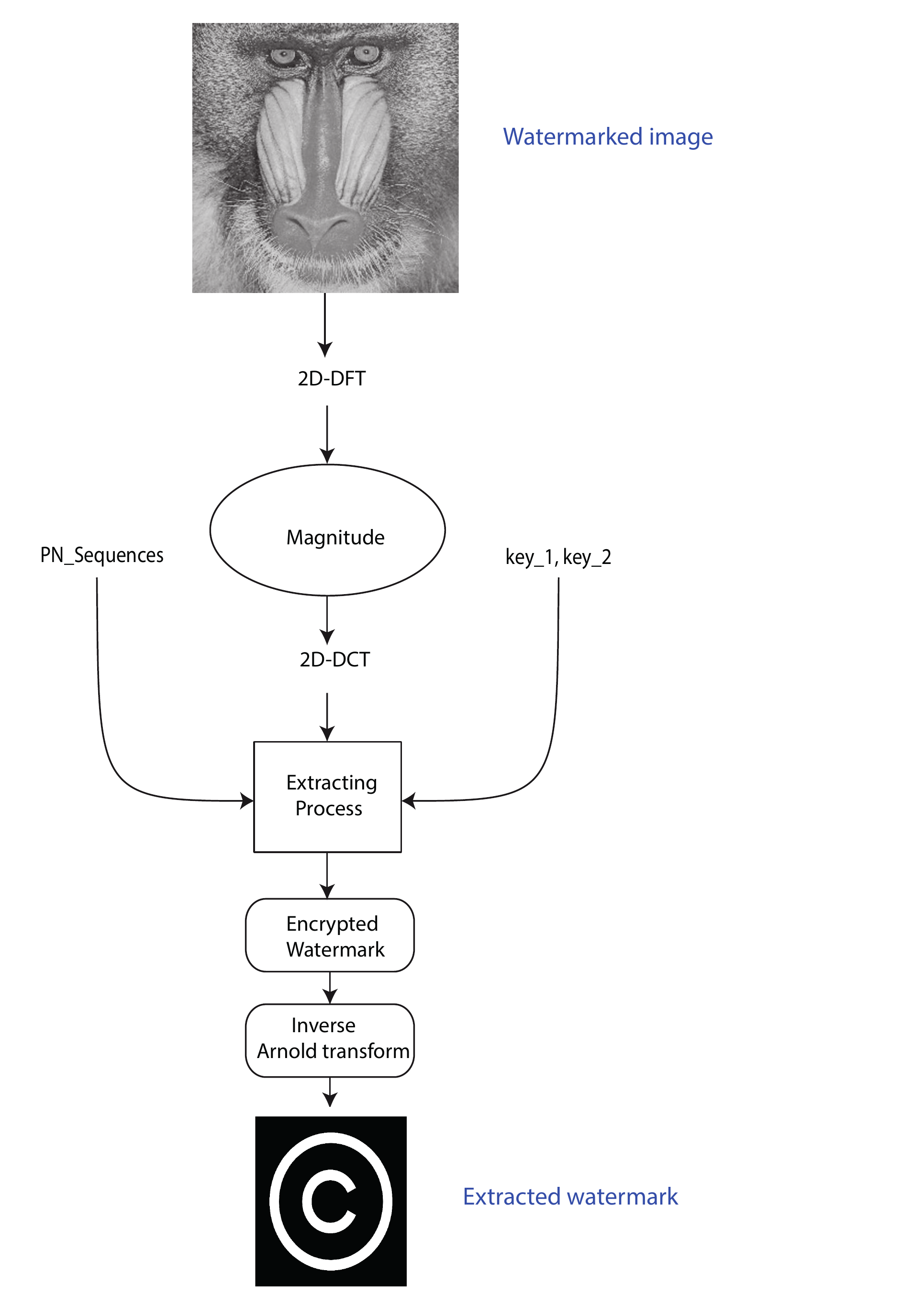}
\caption{The extracting process of the proposed scheme}
\label{fig:Extracting scheme.}
\end{figure}

%As we rely on a blind watermarking scheme, the extraction of the watermark does not need the original image neither the watermark.
%Our  method is a blind watermarking scheme so it doesn't need the original image neither the watermark. 
With the knowledge of the secret key 1 used during the embedding process and the key 2 used during the pretreatment of watermark, the extraction process  is blind and quite simple, as shown in Fig. \ref{fig:Extracting scheme.}. Thus, the proposed method is blind since only  two private keys (key 1 and key 2) are needed.\\
It is sufficient to perform the 2D-DFT of the watermarked image and calculate the  DFT magnitude. With the same secret key (key 1) than in the embedding process  two PN Sequences are generated. Thereby, we obtain the same PN sequences.
 Then, the 2D-DCT is applied to the DFT magnitude. In the extracting process, as shown in Fig. \ref{fig:Extracting scheme.}, the middle-band frequency coefficients of each $8 \times 8$ DCT bloc are extracted. Afterwards, for each bloc, the correlation between the middle band frequencies coefficients and the two PN sequences is computed. Then, the encrypted $i^{th}$ watermark bit is extracted using equation (\ref{eq:Watermarkextraction}). Finally, the inverse Arnold transform using the key 2 is applied to extract the watermark.  The proposed extraction scheme is further described in  Algorithm \ref{alg:WatermarkExtracting}.
\begin{algorithm}[t]
\caption{Watermark extracting}
\label{alg:WatermarkExtracting}
\begin{algorithmic} 
\REQUIRE Watermarked image, key 1, key 2.
\ENSURE  Watermark.
\STATE 1. Apply DFT to the watermarked image and calculate the magnitude.
\STATE 2. Generate two PN sequences (PN\_Seq\_0 and PN\_Seq\_1) using the same secrete key used in the embedding process.  
\STATE 3. Apply DCT to the Magnitude of DFT and extract  the  middle  frequency  band coefficients.
\STATE 4. Calculate  the  correlation  between  the middle frequency band coefficients $F_M$ and the two PN sequences.
\STATE 5. Extract the $i^{th}$ Watermark bit $W_i$ as follows : 
\begin{equation}
\label{eq:Watermarkextraction}
W_i=\begin{Bmatrix}
0 \hspace*{0.5cm}  if \hspace*{0.5cm}  Corr(0) > Corr(1) \\ 
1 \hspace*{0.5cm}  if \hspace*{0.5cm}  Corr(1) > Corr(0) 
\end{Bmatrix}
\end{equation}
Where $Corr(0)$ is  the  correlation  between the middle frequency band coefficients of  $i^{th}$  block and PN\_Seq\_0, and $Corr(1) $  is  the  correlation  between  the middle frequency band coefficients of  $i^{th}$  block and PN\_Seq\_1.
\STATE 6. Perform the inverse Arnold transform and  extract watermark image.
\end{algorithmic}
\end{algorithm}
\section{Experimental setup}
\label{Experimental setup}
%Faut-it ajouter un label à ce chapitre ou pas?
\subsection{Studied images}
\begin{figure}[!h]
\centering
\captionsetup{justification=centering}
%\begin{center}
\includegraphics[width=0.9\textwidth]{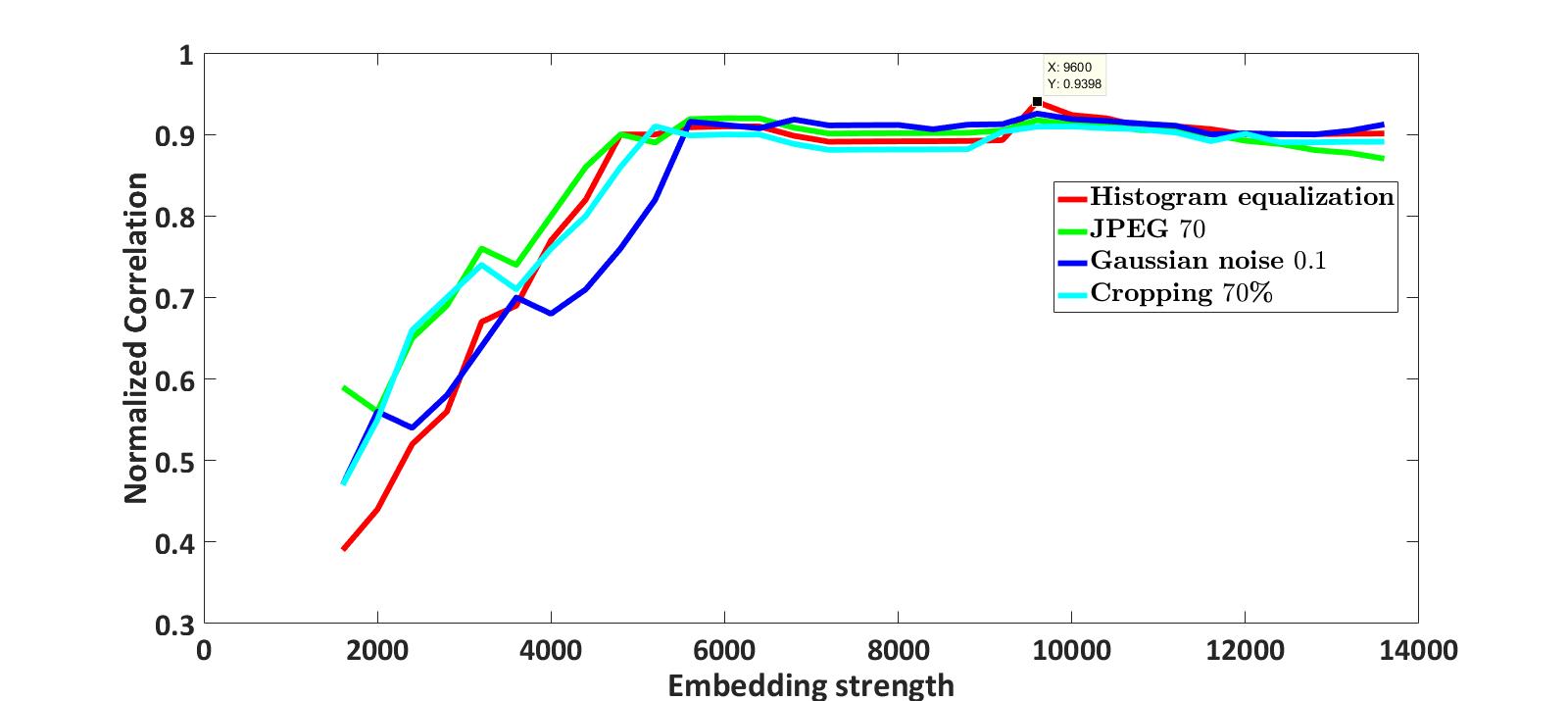}
%\end{center}
\caption{ NC values after variation of embedding strengths under some attacks for Lena}
\label{fig:EmbeddingStrengthWithAttacksLena}
\end{figure}

\begin{figure}[!t]
%\centering
%\captionsetup{justification=centering}
\begin{center}
\includegraphics[width=0.9\textwidth]{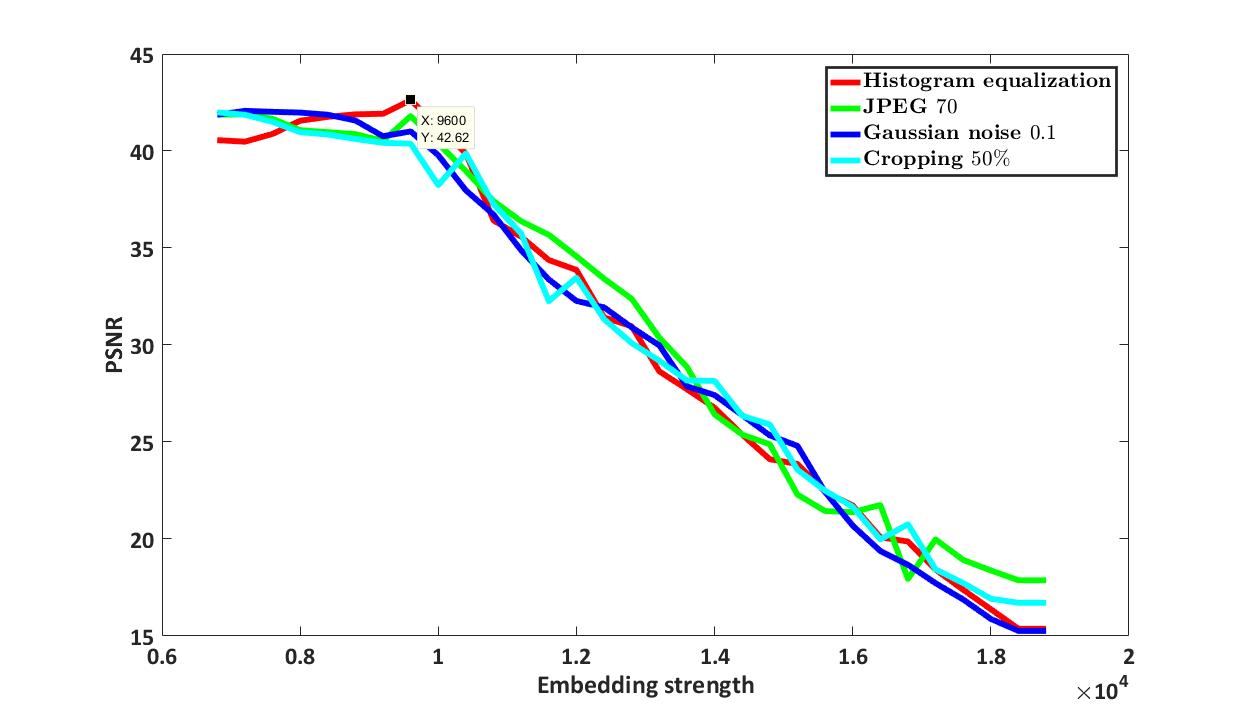}
\end{center}
\caption{ The obtained PSNR values using different embedding strength values after several attacks for Lena image}
\label{fig:EmbeddingStrengthVsPSNRLena}
\end{figure}

To evaluate the performance of the proposed watermarking scheme, several experiments have been conducted on $10$ natural 8-bit grayscale images of size $512\times 512 $  as depicted in Fig. \ref{fig:OriginalAndWatermarkedImages}(a-j) ("Mandril","Peppers","Cameraman", "Lena", "Goldhill", "Walkbridge", "Woman\_blonde", "Livingroom", "Pirate", and "Lake")  and a set of $39$ textured images provided by the University of Southern California \cite{RefJ30}. The majority of these images come from the standard texture image Brodatz database \cite{RefJ31}. Fig. \ref{fig:OriginalAndWatermarkedTexturedImages}(a-j) shows a sample of 10 test textured  images  taken from \cite{RefJ31}. A $(19 \times 52)$  binary logo is used as watermark  as shown in Fig. \ref{fig-ExtractedWatermarksAfterSeveralAttacks} (a).
Another binary logo of size $64\times64$ is used for comparison purpose as show in Fig. \ref{fig:Arnoldtransf.}.
To increase the security and the safety of the watermarking method, the watermark image logo has been scrambled using Arnold transform. 
Parameter value of Arnold transform, which refers to the secret key (key 2), is taken as $n=24$, where $n$ denotes the cycle of Arnold scrambling. The parameter $k$  which denotes the embedding strength of the embedded watermark is chosen in such a way that ensure the best tradeoff between imperceptibility and robustness. To this end, extensive experiments have been conducted using  empirically several values of $k$ to find out the value ensuring this tradeoff. According to these experiments, the best found value is $k=9600$ (see Fig. \ref{fig:EmbeddingStrengthWithAttacksLena} and Fig. \ref{fig:EmbeddingStrengthVsPSNRLena}). Note that this value is the best for the proposed work with or without attack. 
All the experiments were
coded by MATLAB R2013a and implemented on a PC with CPU Intel(R) Core(TM) i5-3470 @ 3.2 GHz with 4-GB of RAM.
\subsection{Evaluation metrics}
%\subsubsection{Imperceptibility}
Numerous metrics have been proposed in the literature to evaluate the quality of images \cite{RefMetric}. When the original image is known, a distance between the original image and the processed one is usually computed. The challenge is to perceptually tune the distance such that the predicted quality is in agreement with human quality judgments. Peak Signal to Noise Ratio (PSNR) is the most widely used metric in the watermarking literature to measure the distance between the original image and the watermarked one.  Althought, it is well recognized that it is does not correlate with human perception we use it for comparative purposes it in this work.
%Metrics used for evaluating image distortion after watermarking are Peak Signal to Noise Ratio (PSNR) and structural similarity (SSIM) index quality.
It is defined as follows : \\
\begin{equation}
PSNR= 10 \log (\frac{MAX^2}{MSE})
\end{equation}
Where MAX is the maximum possible pixel value of the image which is equal to $255$ for an $8-$bit per pixel representation , and MSE is given by : 
\begin{equation}
MSE=\frac{1}{mn} \sum_{i=0}^{m-1}\sum_{j=0}^{n-1} [I(i,j)-K(i,j)]^2
\end{equation}
Where $I(i,j)$ and $K(i,j)$ refers to the original image and the watermarked image respectively. Basically, when the distortions decreases the PSNR increases. \\
%Mettre un tableau qui contient les images et les psnr correspondants avant attaque.
%SSIM 
\\
The structural similarity (SSIM) index performs similarity measurement using a combination of three heuristic factors luminance comparison, contrast comparison, and structure comparison. It is the most influential perceptual quality metric \cite{Refmetric2}. It is defined by \eqref{eq:SSIM}.
\begin{equation}
SSIM(I_0,I_w)=\frac{(2\mu_{I0} \mu_{Iw}+c_1)(2\sigma_{I_0I_W}+c_2)}{(\mu_{I0}^2+ \mu_{Iw}^2+c_1)(\sigma_{I0}^2+\sigma_{Iw}^2+c_2)}
\label{eq:SSIM}
\end{equation}
Where, $I_0$ and $I_w$ are respectively the original image and the watermarked image, 
$\mu_{I0}$ and $\mu_{Iw}$ are respectively the  local means of $I0$ and $Iw$, $\sigma_{I0}^2$ is the variance of $I0$ whereas $\sigma_{Iw}^2$ is the variance of $Iw$,
$c_1$ and $c_2$ are two variables to stabilize the division with weak denominator.
\\
Robustness measure the ability of the watermark to resist against removal due to intentional or unintentional attacks. Indeed, Watermarks should survive standard data processing, such as would occur in a creation and distribution process and also to malicious attack. The normalized correlation (NC) is a widely used attribute for quantifying the robustness of the underlying watermarking technique against various attacks. It measures the similarity between the extracted watermark and the original watermark. It is defined by :
%To evaluate the quality of the extracted watermark, we use the Normalized Correlation (NC), 
\begin{equation}
NC = \frac{ \sum_{i=1}^{M}\sum_{j=1}^{N} \begin{bmatrix}
W(i,j)&\times W'(i,j)
\end{bmatrix}^2}
{\begin{pmatrix}
\sqrt{\sum_{i=1}^{P} \sum_{j=1}^{Q} \left [ W(i,j) \right ]^2}  &  \sqrt{\sum_{i=1}^{P} \sum_{j=1}^{Q} \left [  W'(i,j)\right ]^2}
\end{pmatrix}}
\end{equation}
Where, $W$ and $W'$ are the original and the extracted watermarks, respectively. \\
%An alternative way, that can be find in the literature, in order to measure the robustness is to compute the PSNR between the original and the extracted watermark. 

%Figure of varing the embedding strength explaining the best found value of k

%Embedding strength  Vs PSNR after various attacks
%\begin{figure}[!h]
%%\centering
%\captionsetup{justification=centering}
%\begin{center}
%%\includegraphics[width=1.45\textwidth]{figs/embeddingStrnVSattack.eps}
%\includegraphics[width=1.25\textwidth]{figs/EmbeddingStrenVsAttacks}
%\end{center}
%\caption{NC values under several embedding strengths after several attacks for Lena}
%\label{fig:EmbeddingStrengthAttackedMandrill}
%\end{figure}
%Based on the above experiments, this tradeoff is achieved when $k$ is fixed to $9600$.
%The parameter $k$ denotes the embedding strength of the embedded watermark. Based on the above experiments, the tradeoff between imperceptibility and robustness is achieved when $k$ is set within the value rang $[5500,12500]$.
%Ou je dis directement is achieved when k is fixed to 9600. 
%has been chosen empirically in order to guarantee the tradeoff between imperceptibility and robustness. 
%To this end, we chose, in all our experiments, a watermark strength
%$k = 9600$.
Several common attacks have been applied to these images in order to evaluate the performance of the proposed watermarking algorithm in terms of robustness and imperceptibility.
 %\textcolor{blue}{
 Furthermore, we choose to compare our scheme  with  schemes presented in 
 %\cite{lien2006watermarking},  
 \cite{RefJ14}, \cite{RefJ15}, and  \cite{RefJ16} in terms of  imperceptibility and robustness because they provide a clear presentation and description of their experimental results. 
 %}
%  For this purpose, another $(64 \times 64)$ binary image is used as watermark. Furthermore, to show the robustness of our method in comparison to  
%% \cite{lien2006watermarking}, 
% \cite{wang2004wavelet}, \cite{sahraee2013robust} and \cite{lin2009wavelet}  the NC computed for Lena test image under different attacks by four different approaches is  depicted in Table \ref{tab:Robustnesscomparisonforlena}.
\section{Experimental Results}
\label{Experimental Results}
%\subsection{Evaluation of the proposed method}
\subsection{Imperceptibility}
~\\
In order to evaluate the imperceptibility of the proposed scheme, we calculate the PSNR and the SSIM between the original image and the watermarked image, respectively. Moreover, the absolute difference between watermarked images and original images has been calculated for all test images. For the brevity of space we have given only two, corresponding to the images "Mandrill" and "D94".

\begin{table}[H]
%\footnotesize
\centering
%{\renewcommand{\arraystretch}{1.2}
%\captionsetup{justification=centering}
\caption{Watermark imperceptibility measured in terms of PSNR (dB) and SSIM}
\label{tab:imperceptibility1}
%R??sultats pour k=750
\begin{tabular}{cccccc}
\hline \noalign{\smallskip}
Natural images & \: PSNR \: & \: SSIM \: & \: Textured images  \: & PSNR & \: SSIM \: \\
 \noalign{\smallskip}\hline\noalign{\smallskip}
Mandrill & 61.28& 1.0 & D9 & 58.97 & 1.0 \\

Lena & 61.97&0.9998 & D12 &58.11&1.0 \\

Peppers  &65.97& 1.0&  D94 & 58.26&0.9999 \\ 
\: Cameraman \:  &63.54& 0.9999 &D15 &58.18 & 1.0 \\

\: Goldhill \: &66.37 & 0.9999 & D24 &57.95&1.0 \\

\: Walkbridge \: &59.24&0.9999&  D29 &58.03& 1.0 \\

\: Woman\_blonde \:  &57.31& 0.9998& D38 &57.00&0.9998 \\

\: Livingroom \: &59.37& 0.9999& D84 &57.96&1.0 \\

\: Pirate \:  &58.82&0.9999 & D19&58.38 &0.9999 \\

\: Lake \: & 58.67 &1.0& D112 &58.41&1.0 \\

\: Average\: & 61.25 &0.99991& Average &58.12&0.99996 \\
\noalign{\smallskip} \hline 
\end{tabular}
\\
\end{table}

\begin{figure}[H]
    \centering
    \subfigure[]{\label{sub1} \includegraphics[width=0.38\textwidth]{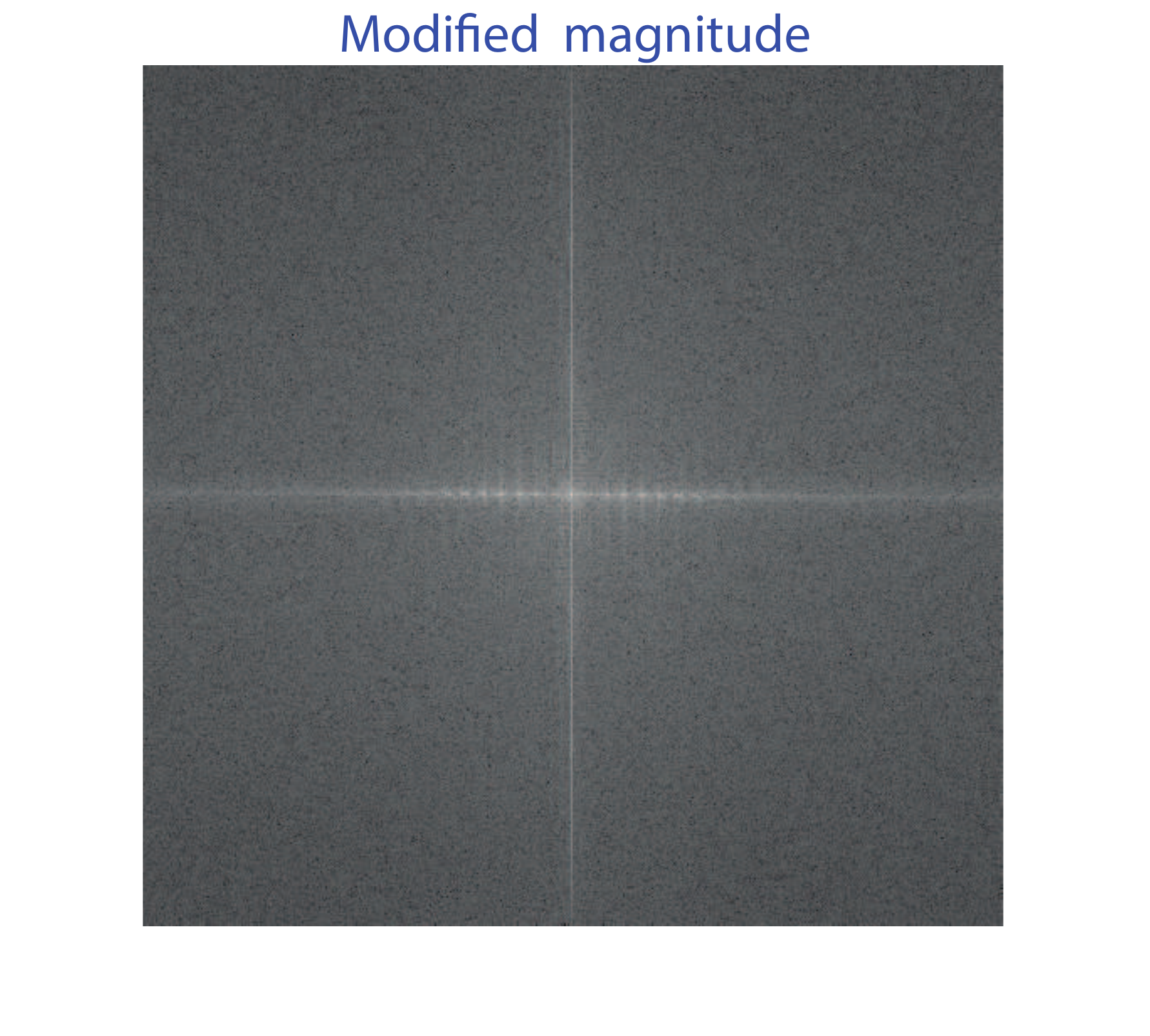}}
    \subfigure[]{\label{sub2} \includegraphics[width=0.38\textwidth]{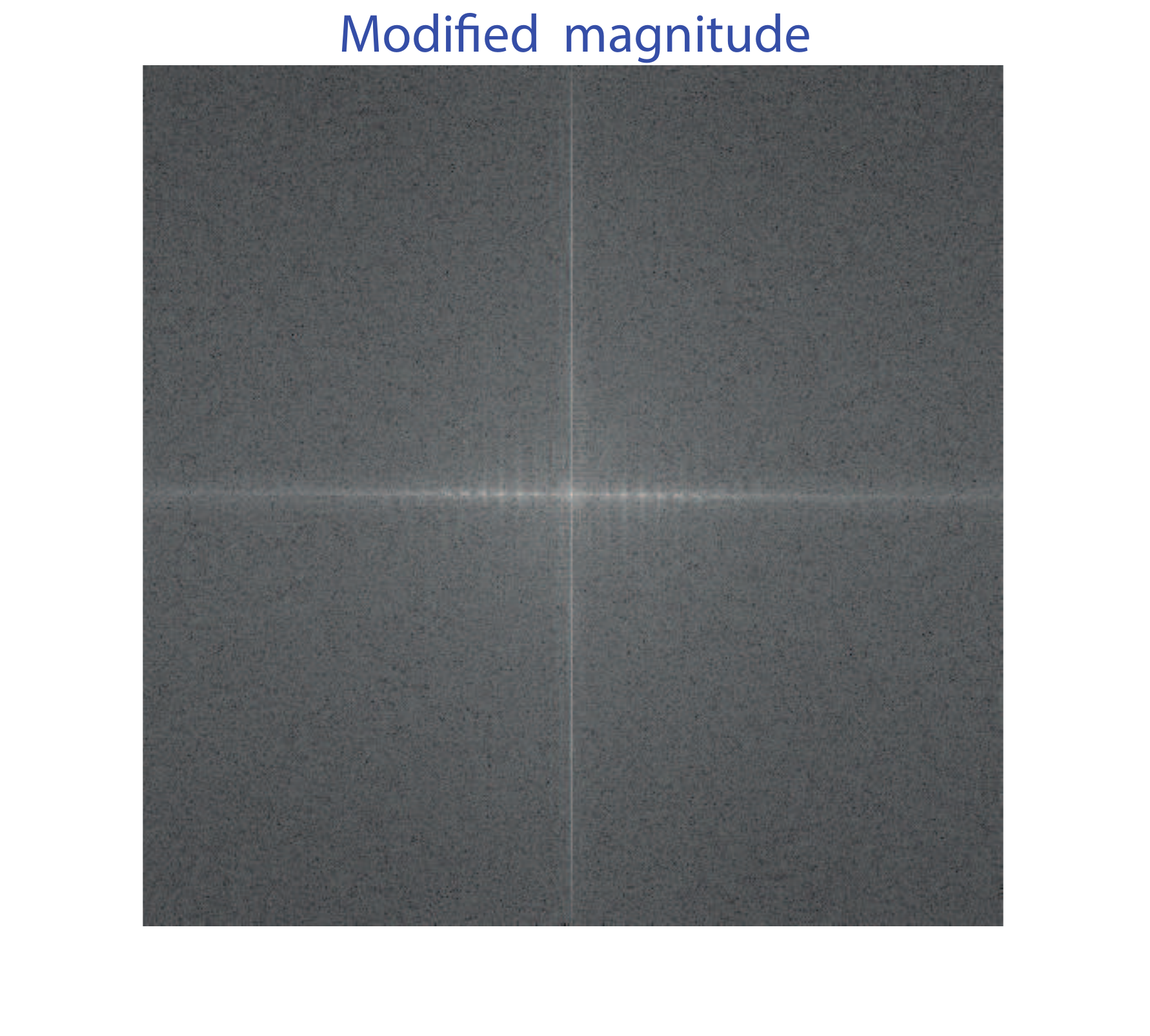}}
\caption{ (a) Original DFT magnitude (b) modified DFT magnitude of Mandrill with ($k=9600$)}
\label{fig:OriginalAndModifiedMagnitude}
\end{figure}
\begin{figure}[H]
    \centering
    \subfigure[]{\label{sub1} \includegraphics[width=0.12\textwidth]{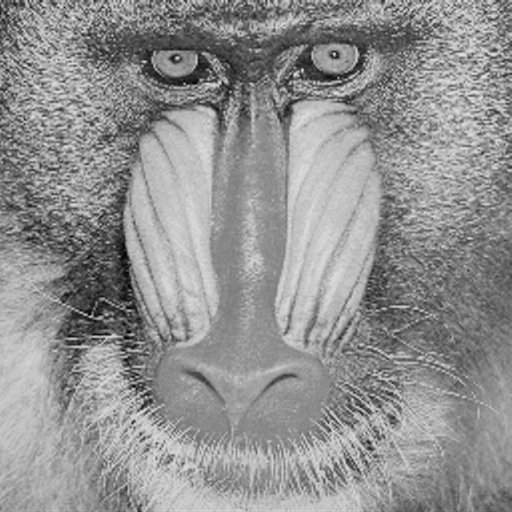}}
    \subfigure[]{\label{sub2} \includegraphics[width=0.12\textwidth]{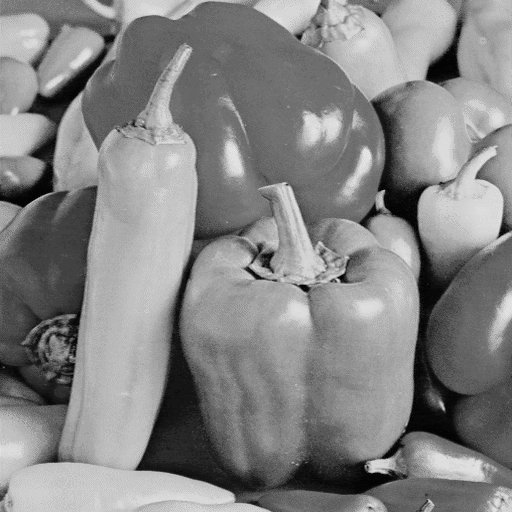}}
    \subfigure[]{\label{sub3} \includegraphics[width=0.12\textwidth]{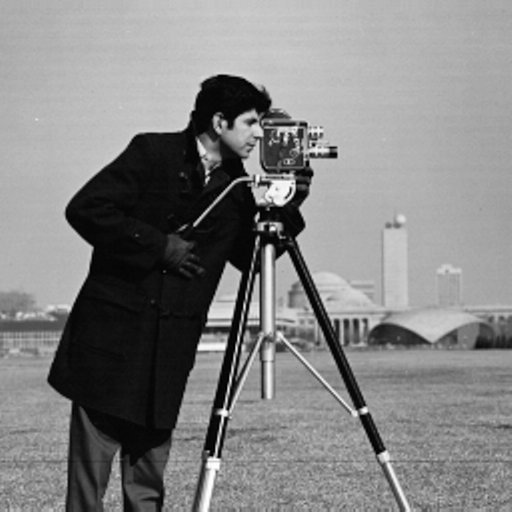}}
    \subfigure[]{\label{sub4} \includegraphics[width=0.12\textwidth]{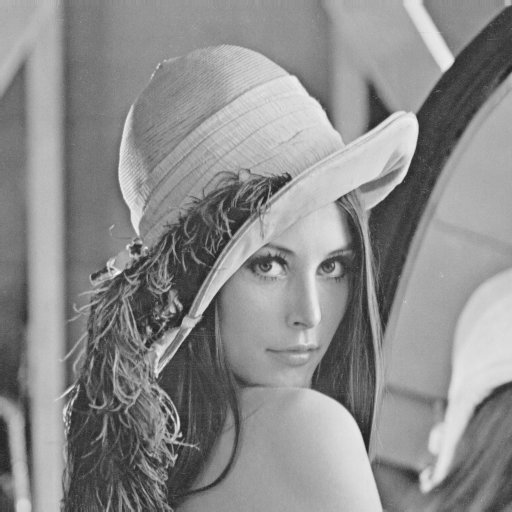}}
\subfigure[]{\label{sub5} \includegraphics[width=0.12\textwidth]{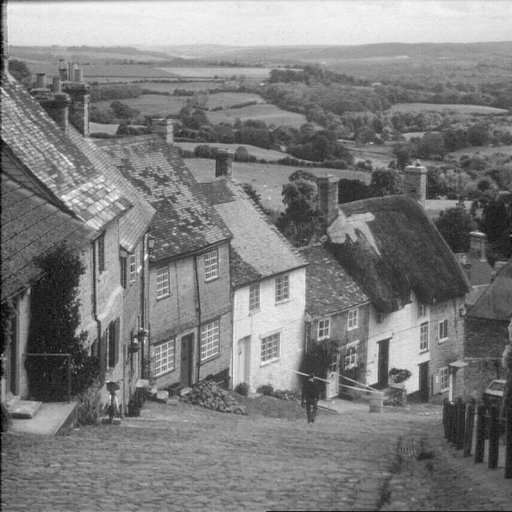}}  

    \subfigure[]{\label{sub6} \includegraphics[width=0.12\textwidth]{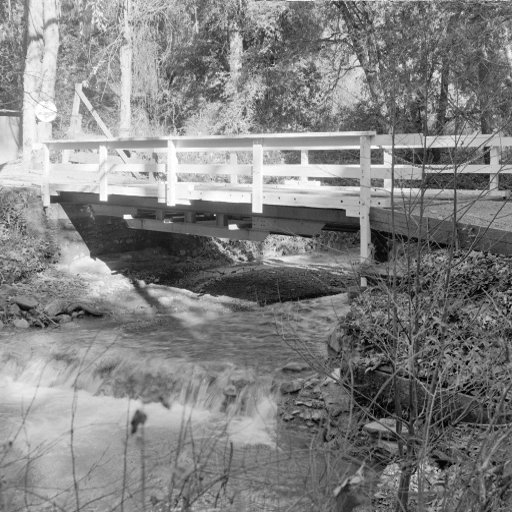}}
    \subfigure[]{\label{sub7} \includegraphics[width=0.12\textwidth]{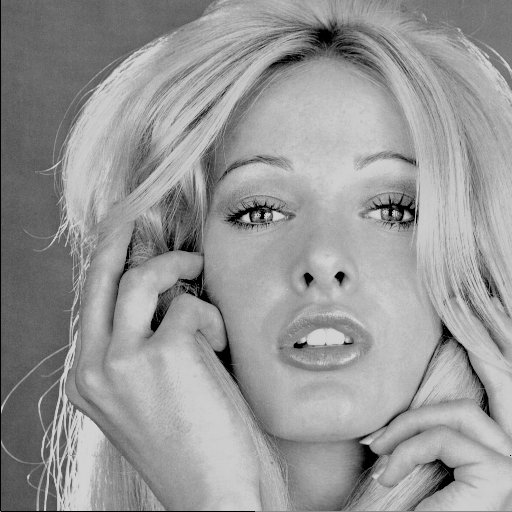}}
    \subfigure[]{\label{sub8} \includegraphics[width=0.12\textwidth]{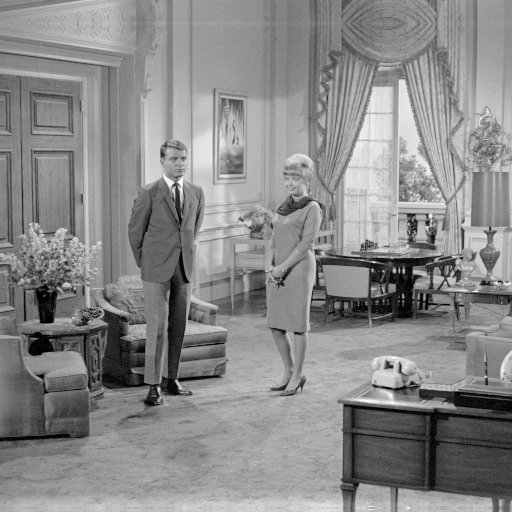}}
    \subfigure[]{\label{sub9} \includegraphics[width=0.12\textwidth]{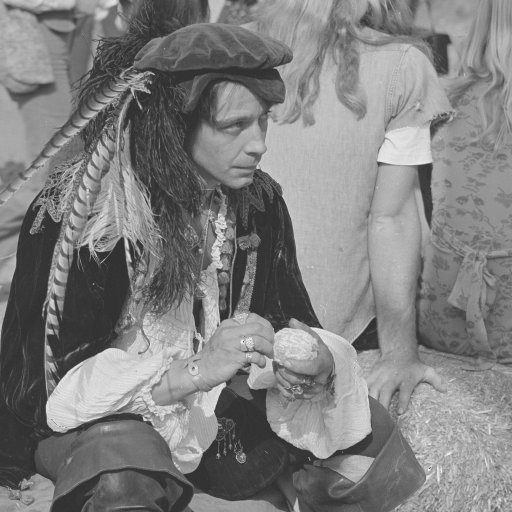}}
\subfigure[]{\label{sub10} \includegraphics[width=0.12\textwidth]{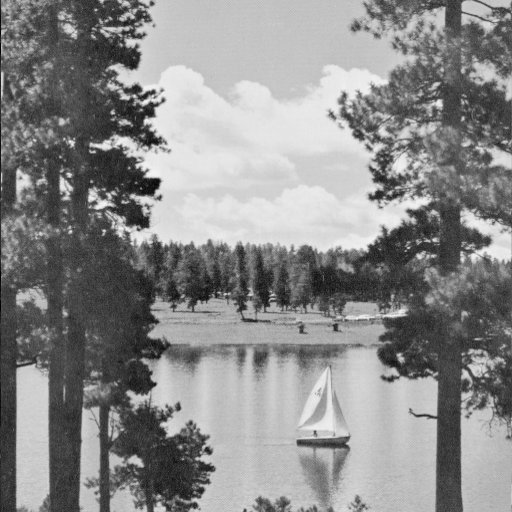}}

\subfigure[]{\label{sub11} \includegraphics[width=0.12\textwidth]{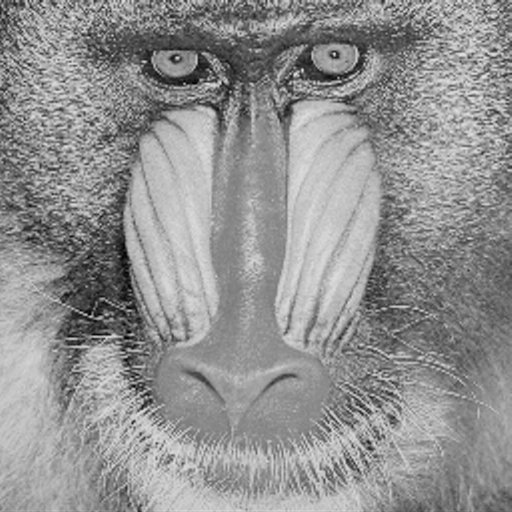}}
    \subfigure[]{\label{sub12} \includegraphics[width=0.12\textwidth]{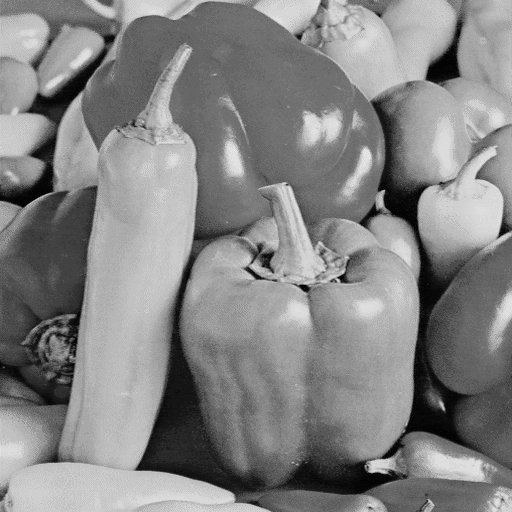}}
    \subfigure[]{\label{sub13} \includegraphics[width=0.12\textwidth]{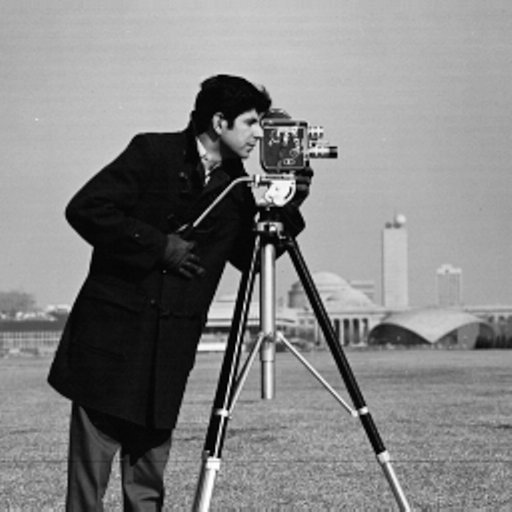}}
    \subfigure[]{\label{sub14} \includegraphics[width=0.12\textwidth]{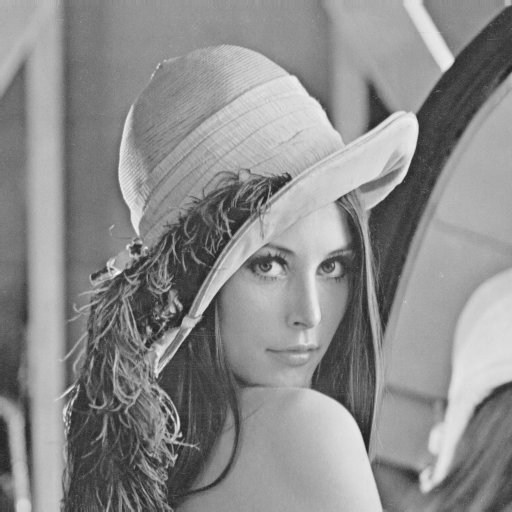}}
\subfigure[]{\label{sub15} \includegraphics[width=0.12\textwidth]{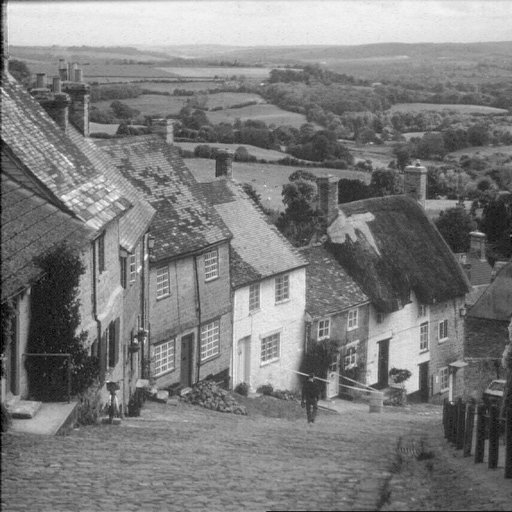}} 

 \subfigure[]{\label{sub16} \includegraphics[width=0.12\textwidth]{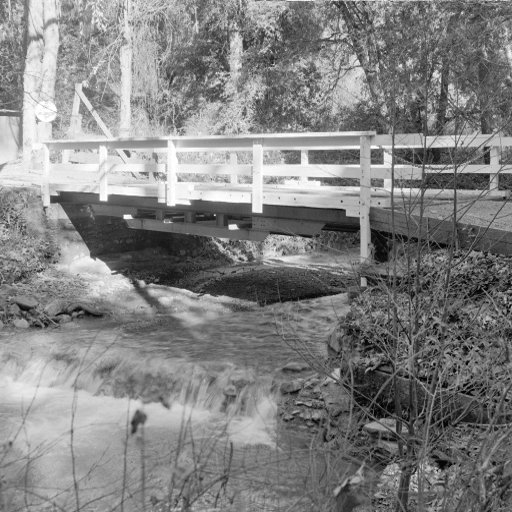}}
    \subfigure[]{\label{sub17} \includegraphics[width=0.12\textwidth]{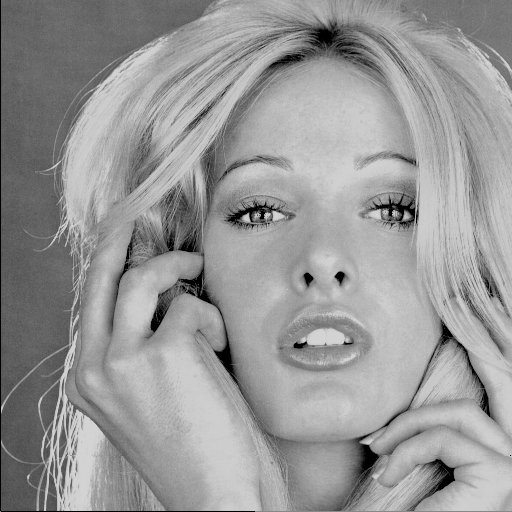}}
    \subfigure[]{\label{sub18} \includegraphics[width=0.12\textwidth]{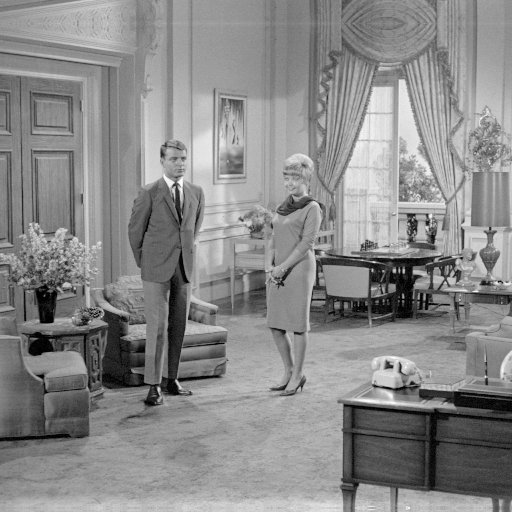}}
    \subfigure[]{\label{sub19} \includegraphics[width=0.12\textwidth]{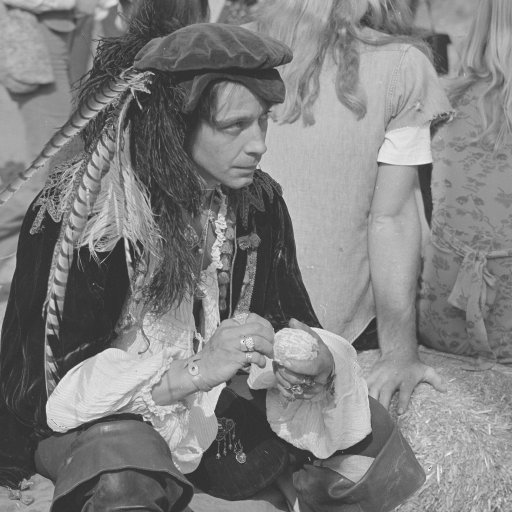}}
\subfigure[]{\label{sub20} \includegraphics[width=0.12\textwidth]{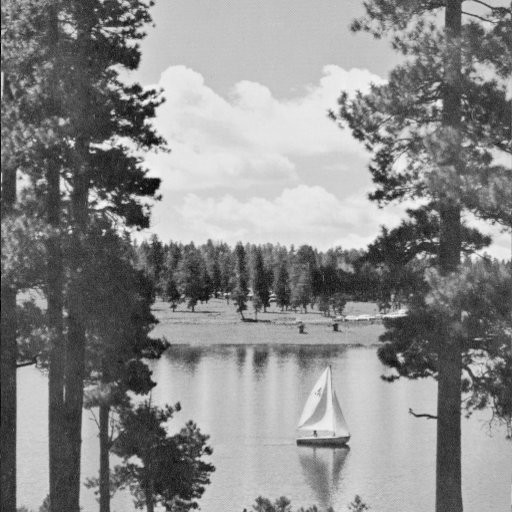}}
    \caption{Original images: (a) Mandril, (b) Peppers, (c) Cameraman, (d) Lena, (e) Goldhill, (f) Watermarked Mandril, (g) Watermarked Peppers,(h) Watermarked Cameraman, (i) Watermarked Lena, (j) Watermarked Goldhill}
    \label{fig:OriginalAndWatermarkedImages}
\end{figure}

\begin{figure}[!]
    \centering
%    \subfigure[]{\label{sub111} \includegraphics[width=0.72\textwidth]{figs/diff1.eps}}
%    \subfigure[]{\label{sub121} \includegraphics[width=0.2\textwidth]{figs/difference_peppers_k_750.eps}}
   \subfigure[]{\label{sub98} \includegraphics[width=0.486\textwidth]{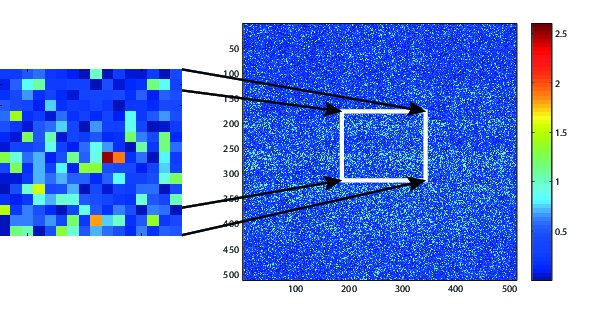}}
%    \subfigure[]{\label{sub141} \includegraphics[width=0.2\textwidth]{figs/difference_lena_k_750.eps}}
%\subfigure[]{\label{sub151} \includegraphics[width=0.2\textwidth]{figs/difference_goldhill_k_750.eps}}  
%\subfigure[]{\label{sub161} \includegraphics[width=0.2\textwidth]{figs/difference_walkbridge_k_750.eps}}
%    \subfigure[]{\label{sub171} \includegraphics[width=0.2\textwidth]{figs/difference_women_blonde_k_750.eps}}
%    \subfigure[]{\label{sub181} \includegraphics[width=0.2\textwidth]{figs/difference_livingroom_k_750.eps}}
    \subfigure[]{\label{sub99} \includegraphics[width=0.486\textwidth]{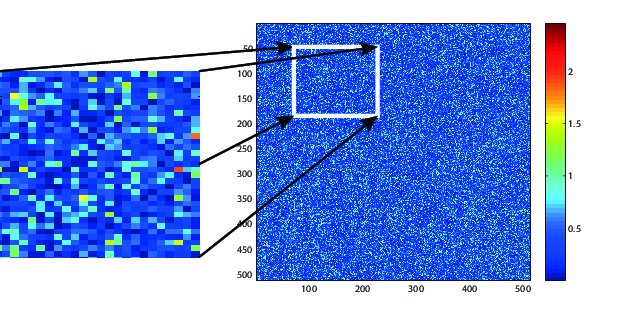}}
%\subfigure[]{\label{sub201} \includegraphics[width=0.2\textwidth]{figs/difference_lake_k_750.eps}}
\caption{Absolute difference of original image and watermarked image : (a) Mandrill, (b) D94}
\label{fig:DifferenceofNaturelimages}
\end{figure}

\begin{figure}[H]
    \centering
    \subfigure[]{\label{sub11} \includegraphics[width=0.17\textwidth]{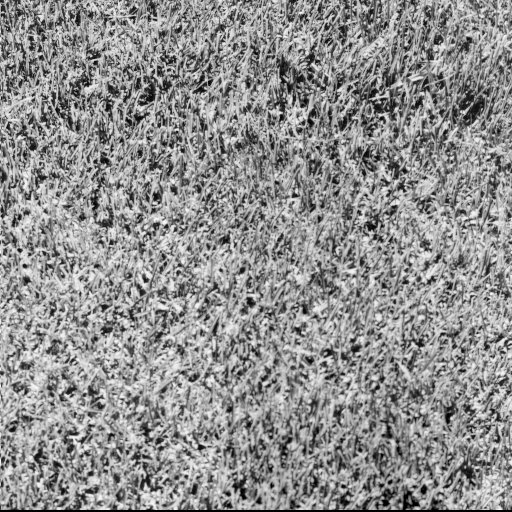}}
    \subfigure[]{\label{sub12} \includegraphics[width=0.17\textwidth]{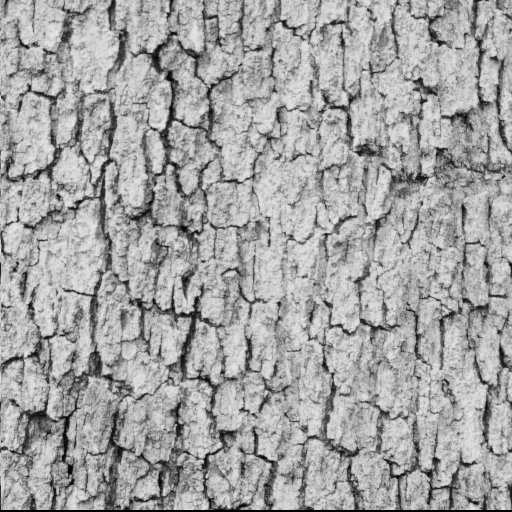}}
    \subfigure[]{\label{sub13} \includegraphics[width=0.17\textwidth]{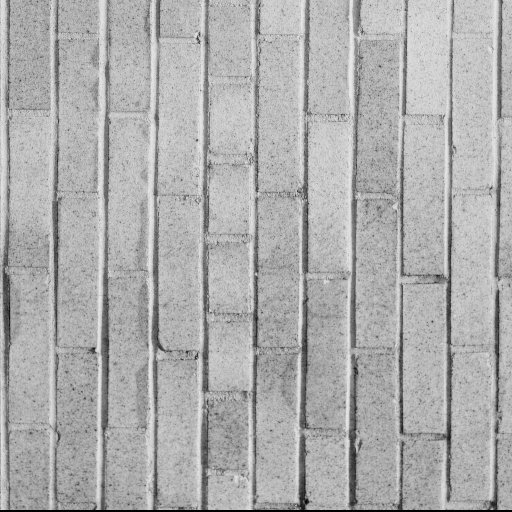}}
    \subfigure[]{\label{sub14} \includegraphics[width=0.17\textwidth]{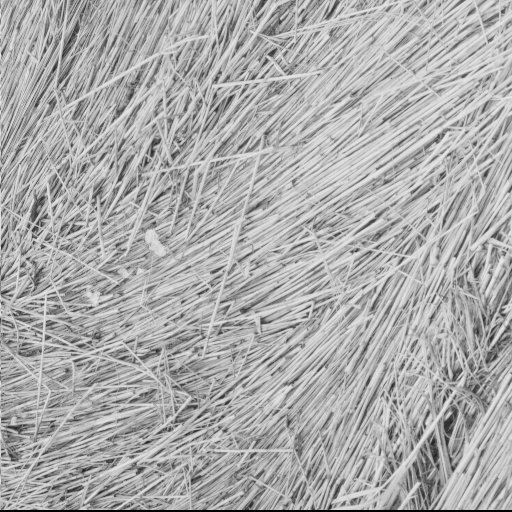}}
\subfigure[]{\label{sub15} \includegraphics[width=0.17\textwidth]{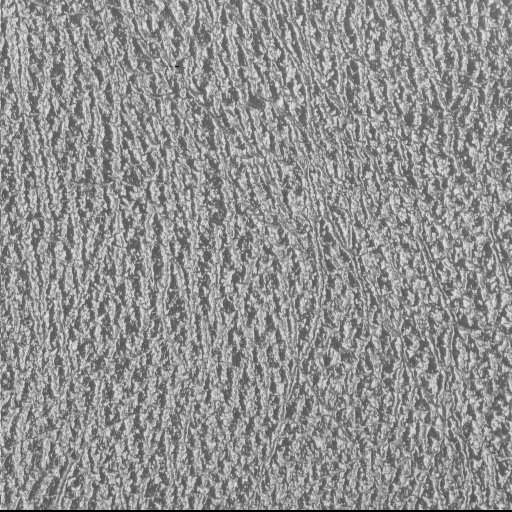}}  
\subfigure[]{\label{sub16} \includegraphics[width=0.17\textwidth]{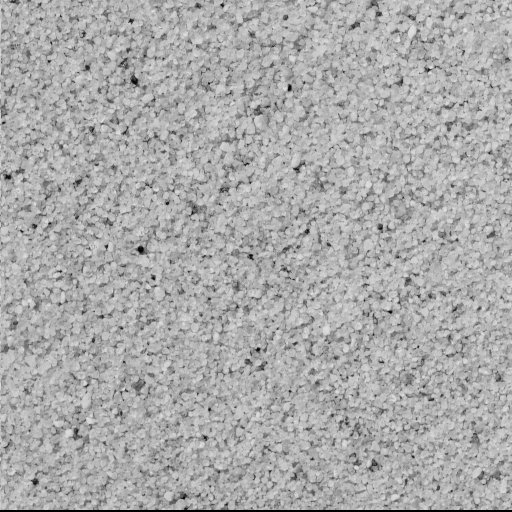}}
    \subfigure[]{\label{sub17} \includegraphics[width=0.17\textwidth]{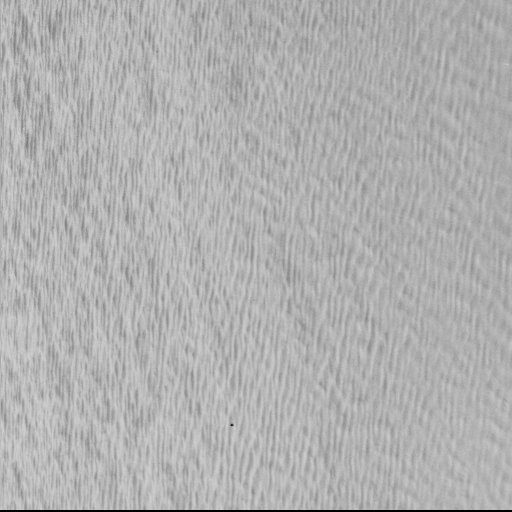}}
    \subfigure[]{\label{sub18} \includegraphics[width=0.17\textwidth]{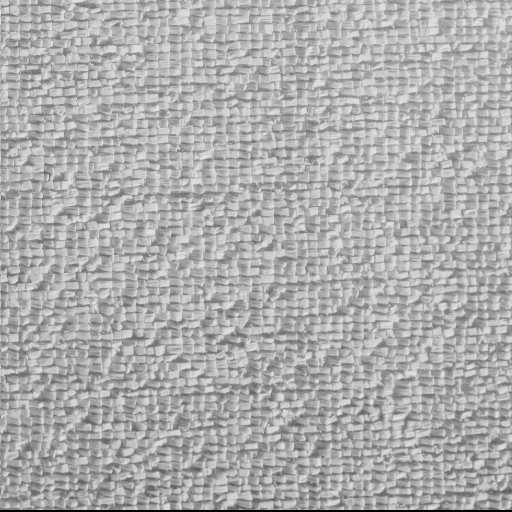}}
    \subfigure[]{\label{sub19} \includegraphics[width=0.17\textwidth]{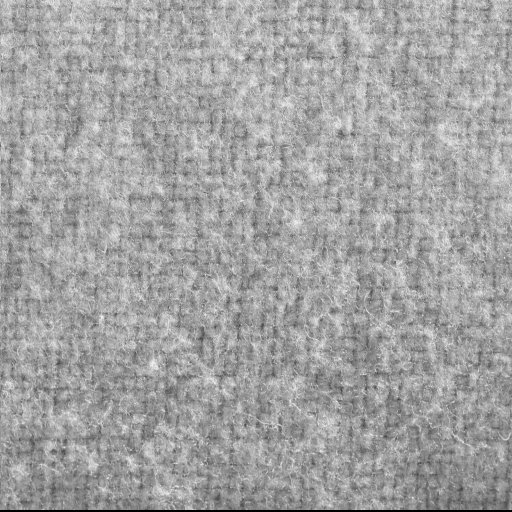}}
\subfigure[]{\label{sub20} \includegraphics[width=0.17\textwidth]{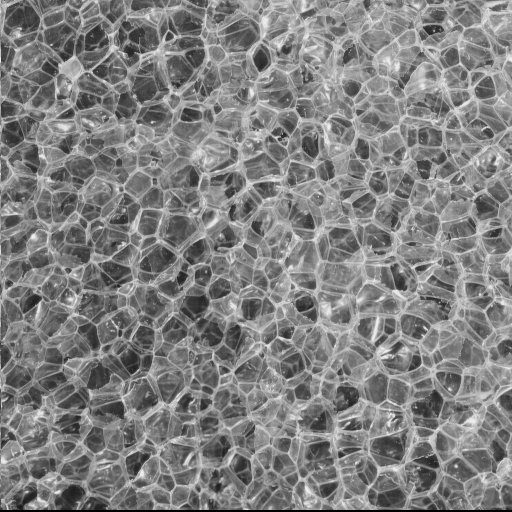}} 

\subfigure[]{\label{sub21} \includegraphics[width=0.17\textwidth]{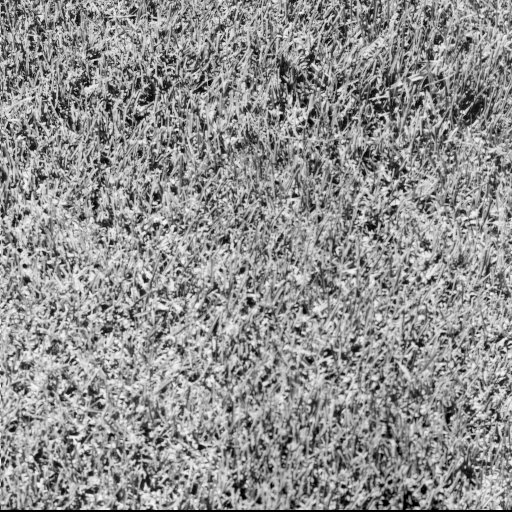}}
    \subfigure[]{\label{sub22} \includegraphics[width=0.17\textwidth]{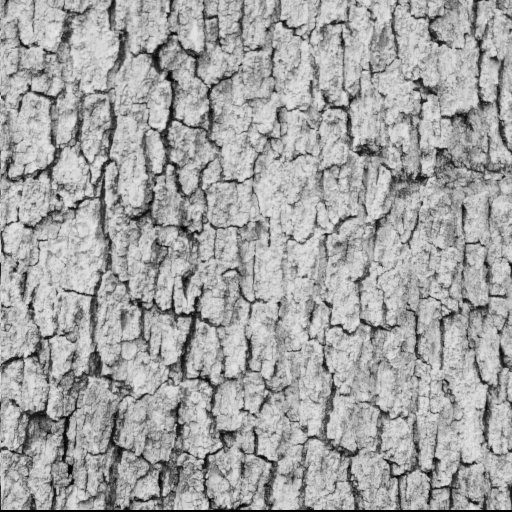}}
    \subfigure[]{\label{sub23} \includegraphics[width=0.17\textwidth]{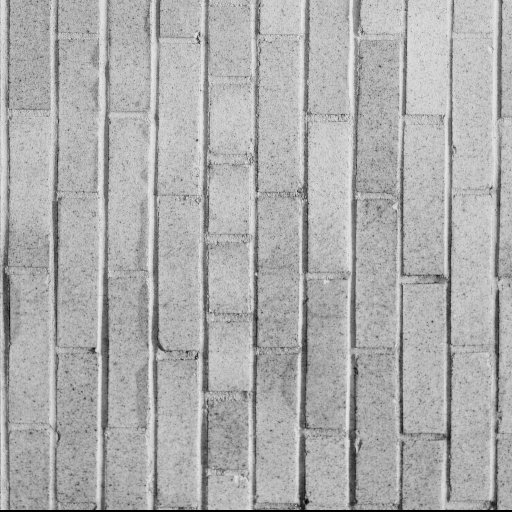}}
    \subfigure[]{\label{sub24} \includegraphics[width=0.17\textwidth]{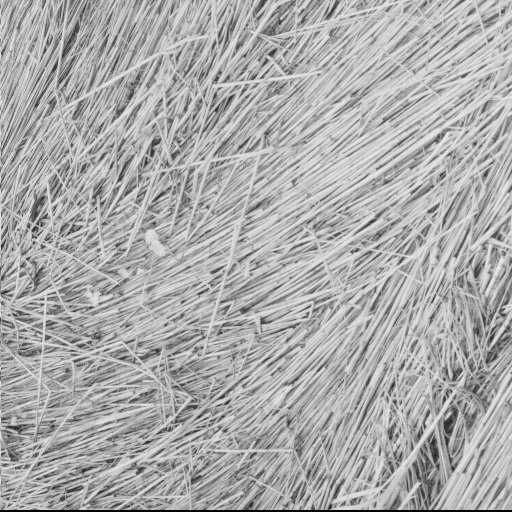}}
\subfigure[]{\label{sub25} \includegraphics[width=0.17\textwidth]{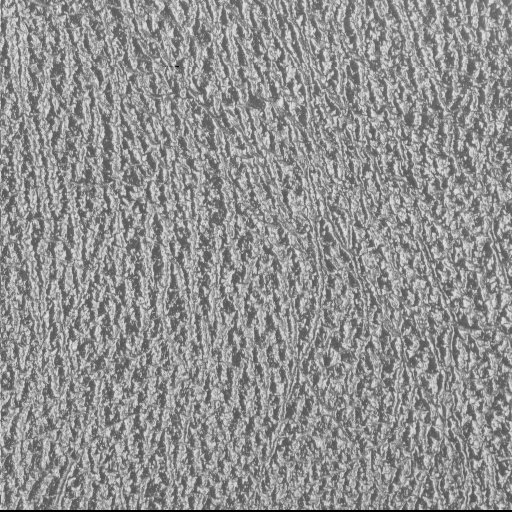}}  
\subfigure[]{\label{sub26} \includegraphics[width=0.17\textwidth]{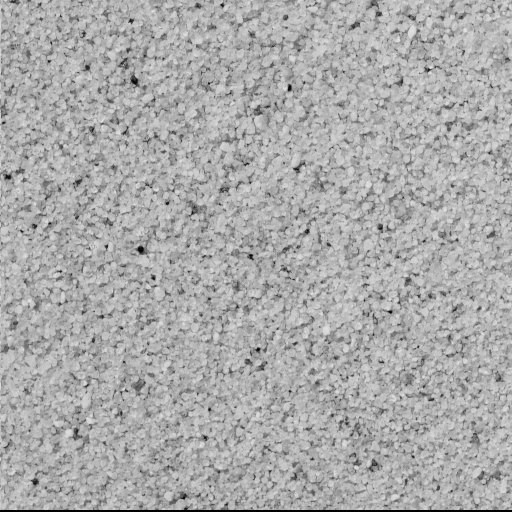}}
    \subfigure[]{\label{sub27} \includegraphics[width=0.17\textwidth]{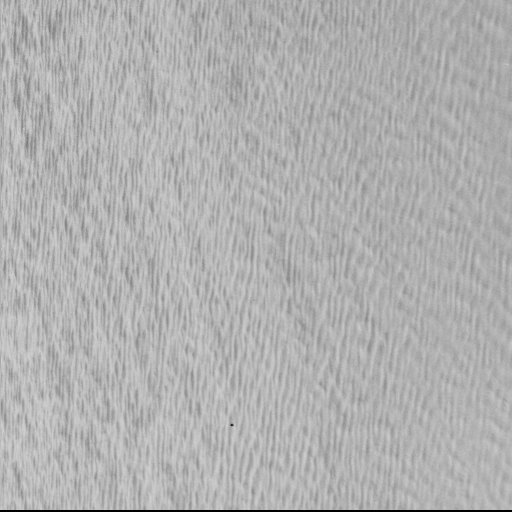}}
    \subfigure[]{\label{sub28} \includegraphics[width=0.17\textwidth]{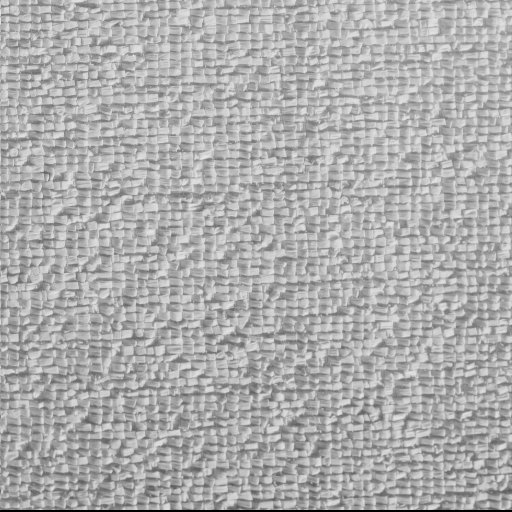}}
    \subfigure[]{\label{sub29} \includegraphics[width=0.17\textwidth]{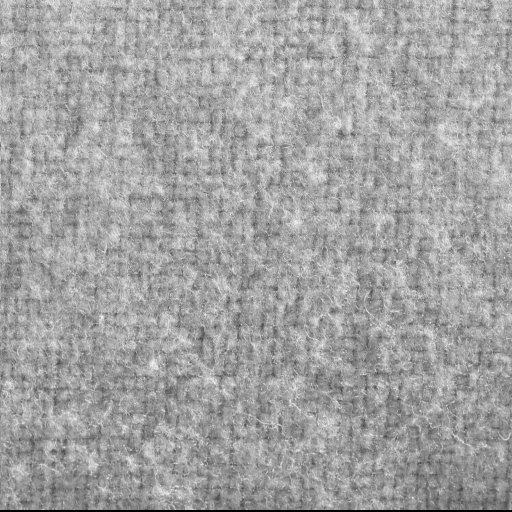}}
\subfigure[]{\label{sub30} \includegraphics[width=0.17\textwidth]{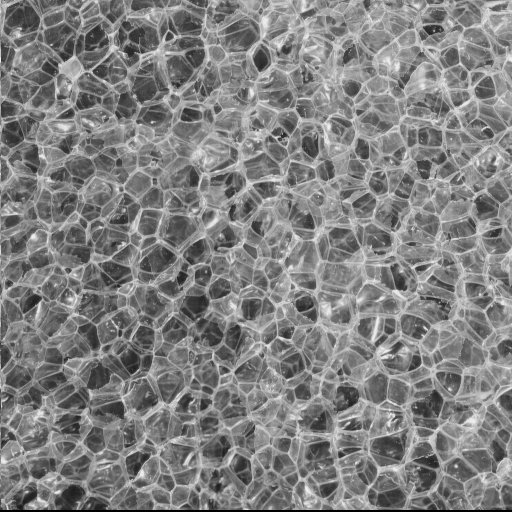}} 

    \caption{Sample of original textured images from Brodatz \cite{RefJ31} : (a-j). Watermarked textured images : (k-t)}
    \label{fig:OriginalAndWatermarkedTexturedImages}
\end{figure}

%\begin{figure}
%    \centering
%%    \subfigure[]{\label{sub111} \includegraphics[width=0.2\textwidth]{figs/textures/difference_textured1.eps}}
%%    \subfigure[]{\label{sub121} \includegraphics[width=0.2\textwidth]{figs/textures/difference_textured2.eps}}
%    \subfigure[]{\label{sub131} \includegraphics[width=0.8\textwidth]{figs/textures/diff_text1.eps}}
%%    \subfigure[]{\label{sub141} \includegraphics[width=0.2\textwidth]{figs/textures/difference_textured4.eps}}
%%	\subfigure[]{\label{sub151} \includegraphics[width=0.2\textwidth]{figs/textures/difference_textured5.eps}}  
%%	\subfigure[]{\label{sub161} \includegraphics[width=0.2\textwidth]{figs/textures/difference_textured6.eps}}
%    \subfigure[]{\label{sub171} \includegraphics[width=	0.8\textwidth]{figs/textures/diff_text2.eps}}
%    \subfigure[]{\label{sub181} \includegraphics[width=0.8\textwidth]{figs/textures/difference_textured8.eps}}
%%    \subfigure[]{\label{sub191} \includegraphics[width=0.2\textwidth]{figs/textures/difference_textured9.eps}}
%%\subfigure[]{\label{sub201} \includegraphics[width=0.2\textwidth]{figs/textures/difference_textured10.eps}}
%\caption{Difference images of 3 samples of textured images taken from Brodatz database \cite{RefJ31}}
%\label{fig:DifferenceofTexturedimages}
%\end{figure}

\begin{figure}[t]
\centering
\subfigure[]{\label{sub200}\includegraphics[width=0.42\textwidth]{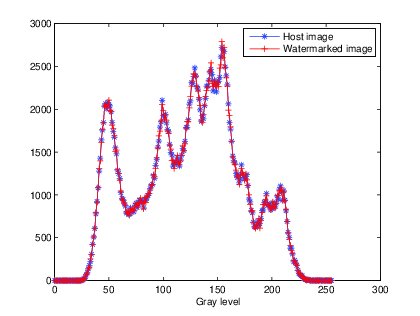}}
\subfigure[]{\label{sub201}\includegraphics[width=0.42\textwidth]{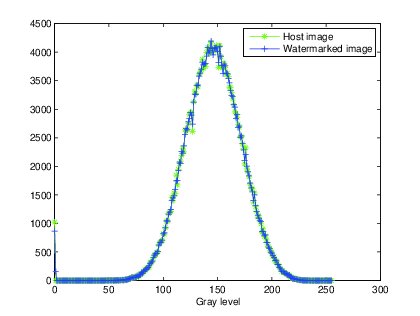}}
\caption{(a) : Imperceptibility illustration through histograms taking image Lena, (b):Imperceptibility illustration through histograms taking image D19 taken from Brodatz }
\label{fig:Imperceptibility_illustration_through_histograms_for_lena.}
\end{figure}
%\begin{figure}[t]
%\centering
%\includegraphics[width=0.40\textwidth]{figs/hist_orig_D19_vs_watermarked_D19.eps}
%\caption{Imperceptibility illustration through histograms taking image D19 taken from Brodatz}
%\label{fig:Imperceptibility_illustration_through_histograms_for_D19.}
%\end{figure}
Table \ref{tab:imperceptibility1} shows the imperceptibility results in terms of  PSNR and SSIM using a $(19 \times 52)$ logo as watermark.
From Fig. \ref{fig:OriginalAndWatermarkedImages} and Table \ref{tab:imperceptibility1}, it can be seen that the watermarked images preserve good visible quality and thus there is no visual distortion. Besides, all the obtained PSNR values are above  $57$ dB and the SSIM values are close to $1$. 
In addition, the PSNR average value of natural images  slightly exceeds the PSNR average of the textured images. These results demonstrate that the proposed method is very insensitive to the image nature.
 \par 
 It can be observed from the difference between the watermarked images and the original images shown in Fig. \ref{fig:DifferenceofNaturelimages} 
% and Fig.\ref{fig:DifferenceofTexturedimages} 
 that the modified part are spread out over the image. This is  due to the fact that the watermark is embedded in all the coefficients of the middle band of DCT. 
It can be concluded from the above figures that all the obtained values  after calculating the difference between original images and watermarked images are close to $0$. In addition, as shown in Fig. \ref{fig:OriginalAndModifiedMagnitude}, there is no visual difference between the original magnitude and the modified one after watermark embedding.
%The histogram is a good way to obtain the tonal distribution of an image. Thus, the viewer can judge the entire tonal distribution at a glance \cite{RefJ32}. 
Furthermore, we have compared the histogram of the original and watermarked images to check if there is any clue that the image has been watermarked. Indeed, in some watermarking techniques the distribution of the watermarked image is unbalanced, suggesting the presence of a watermark.
Due to space limitations, we report only two typical results, corresponding to the images "Lena" and "D19" in Fig. \ref{fig:Imperceptibility_illustration_through_histograms_for_lena.} . We can observe from this Figure  the similarity between the shape of the histograms of the host image and the watermarked image. 
% In conclusion, our proposal shows high imperceptibility regardless of the image type.\\
% \textcolor{blue}{

In order to quantify the impact of using two transforms in the proposed scheme, we compare the performance in terms of PSNR and SSIM between the DFT-DCT and the DFT-only based approach. It can be seen from Table \ref{tab:ImperceptibilitycomparisonforDFTonlyAndDftDct}  that the combination of the  two transforms DFT-DCT gives better results in terms of imperceptibility than the DFT-only based approach for all the test images.
%}
 \begin{table}[H]
%\footnotesize
\centering
%{\renewcommand{\arraystretch}{1.9}
%{\setlength{\tabcolsep}{0.7cm} 
\caption{Comparison of the imperceptibility between the DFT only and the  DFT-DCT  based algorithm for several images}
\label{tab:ImperceptibilitycomparisonforDFTonlyAndDftDct}
\begin{tabular}{ccccc}
\hline \noalign{\smallskip} 
% & \multicolumn{2}{c|}  {Images}     \tabularnewline
%\cline{2-3}
 & \multicolumn{4}{c}  {Watermarking methods}     \\
\cline{2-5}
  \: Cover image \: &  DFT  only   &  &DFT-DCT& \\ 
\hline

 & \multicolumn{4}{c}  {Imperceptibility metric}   \\
\cline{2-5}
  & \: PSNR \:  & SSIM &PSNR& SSIM\\

%No Attack  & Infinity & \textbf{1.0} &&&&\tabularnewline  
%\hline
Mandrill &45.58& 0.9871&61.28& 1.0 \\
Lena  & 47.31 & 0.9785&61.97&0.9998\\
Peppers & 48.21 & 0.9745 &65.97&1.0\\
Cameraman & 46.15  &0.9803&63.54&0.9999\\
Goldhill &  49.73 &0.9762&66.37&0.9999\\ 
Walkbridge & 41.02  &0.9778 &59.24&0.9999\\  
Woman\_blonde  &39.87&0.9812 &57.31&0.9998\\
Livingroom& 43.66   & 0.9788 &59.37&0.9999\\
Pirate& 44.81  &0.9801&58.82&0.9999\\
Lake & 44.05&0.9796 &58.67&1.0\\
Average &45.039 &0.9794&61.25&0.9991\\
\hline  
D9 & 44.37 & 0.9692&58.97& 1.0 \\
D12  & 44.63&0.9655 &58.11&1.0\\
D94 & 44.58  & 0.9745 &58.26&0.9999\\
D15 & 44.66  &0.9758&58.18&1.0\\
D24 &  43.92 &0.9678&57.95&1.0\\ 
D29 & 43.98 & 0.9754 &58.03&1.0\\  
D38  &43.45 & 0.9620&57.00&0.9998\\
D84& 44.01   & 0.9685 &57.96&1.0\\
D19& 44.85  &0.9588&58.38&0.9999\\
D112 & 45.02  &0.9734 &58.41&1.0\\
Average & 44.347 &0.9691& 58.125&0.9996\\
\noalign{\smallskip}\hline
\end{tabular}
\end{table}

%}
\begin{figure}[!t]
\centering
\includegraphics[width=0.70\textwidth]{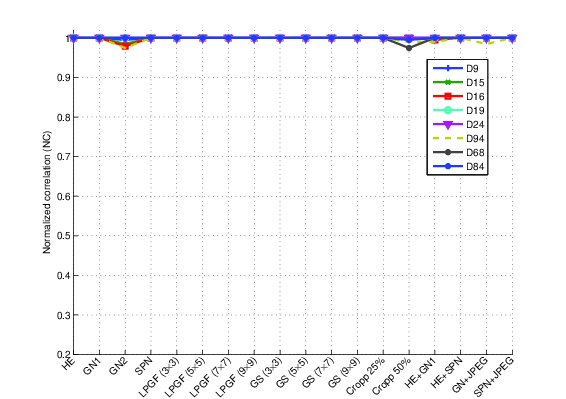}
\caption{Robustness in terms of NC after several attacks applied to a simple of textured images}
\label{fig:Robustness_in_terms_of_NC_after_several_attacks_applied_to_a_simple_of_textured_images}
\end{figure}

\begin{figure}[!t]
\centering
\includegraphics[width=0.75\textwidth]{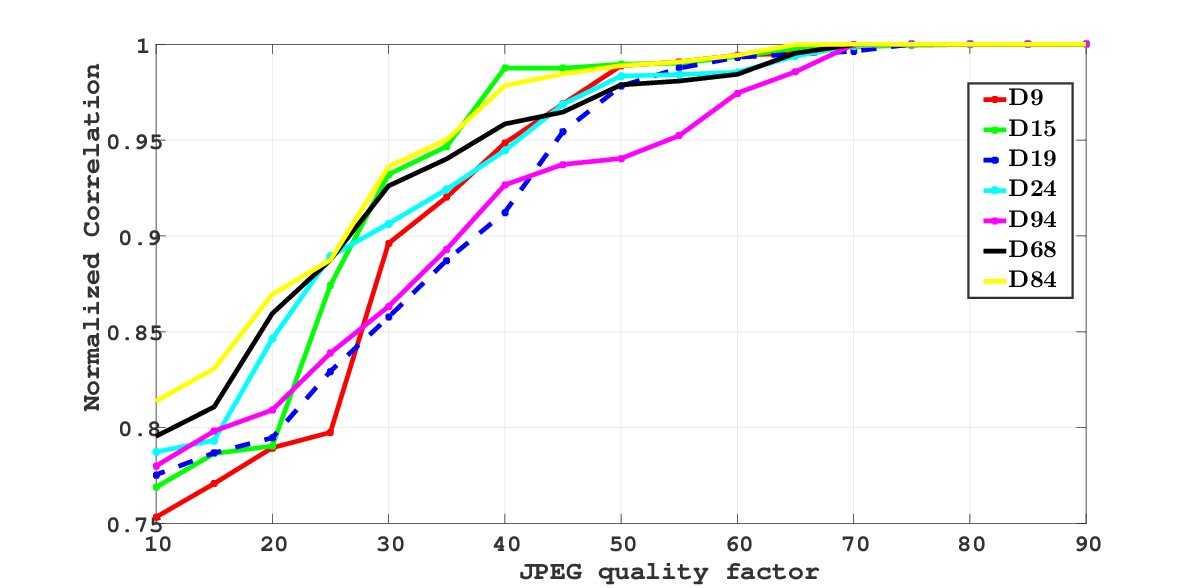}
\caption{Robustness in terms of NC after JPEG compression attack applied to a simple of textured image}
\label{fig:Robustness in terms of NC after JPEG compression attack applied to a simple of textured image.}
\end{figure}

\begin{figure}
\centering
\includegraphics[width=0.65\textwidth]{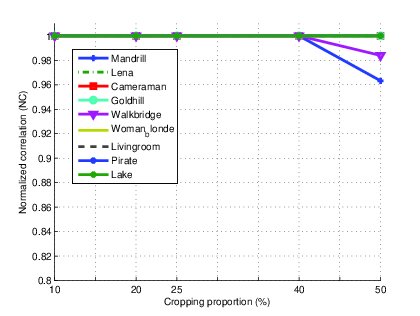}
\caption{Robustness in terms of NC after cropping attack applied to a simple of natural images}
\label{fig:Robustness in terms of NC after cropping attack applied to a simple of natural image.}
\end{figure}

\begin{figure}
\centering
\includegraphics[width=0.75\textwidth]{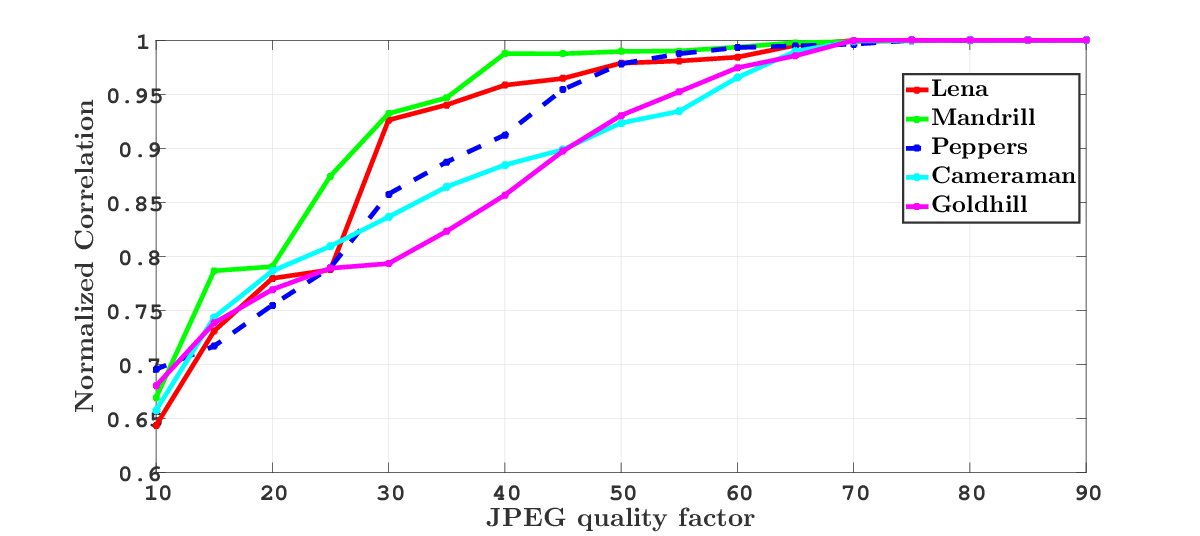}
\caption{Robustness in terms of NC after JPEG compression attack applied to a simple of natural images}
\label{fig:Robustness in terms of NC after JPEG compression attack applied to a simple of natural image.}
\end{figure}
%\subsubsection{Robustness}
~\\
% Robustness measure the ability of the watermark to resist against removal due to intentional or unintentional attacks. Indeed, Watermarks should survive standard data processing, such as would occur in a creation and distribution process and also to malicious attack. The normalized correlation (NC) is a widely used attribute for quantifying the robustness of the underlying watermarking technique against various attacks. It measures the similarity between the extracted watermark and the original watermark. It is defined by :
%%To evaluate the quality of the extracted watermark, we use the Normalized Correlation (NC), 
%\begin{equation}
%NC = \frac{ \sum_{i=1}^{M}\sum_{j=1}^{N} \begin{bmatrix}
%W(i,j)&\times W'(i,j)
%\end{bmatrix}^2}
%{\begin{pmatrix}
%\sqrt{\sum_{i=1}^{P} \sum_{j=1}^{Q} \left [ W(i,j) \right ]^2}  &  \sqrt{\sum_{i=1}^{P} \sum_{j=1}^{Q} \left [  W'(i,j)\right ]^2}
%\end{pmatrix}}
%\end{equation}
%Where, $W$ and $W'$ are the original and the extracted watermarks, respectively. \\
%An alternative way, that can be find in the literature, in order to measure the robustness is to compute the PSNR between the original and the extracted watermark. 
%\\
\subsection{Robustness}
\begin{table}[!]
%\footnotesize
%\centering
%{\renewcommand{\arraystretch}{1.2} %donne la distance entre les lignes%
%{\setlength{\tabcolsep}{0.6cm} %donne la distance entre les collones%
\caption{SSIM and NC values after several attacks applied to D9 image taken from the Brodatz database \cite{RefJ31}}
\label{tab:SSIM_and_NC_value_after_several_attacks_TexturedImg.}
%R??sultats pour l'image baboon
\begin{tabular}{ccc}

\hline \noalign{\smallskip}
\: Attacks \:  & SSIM & \: NC \: \\
 
\hline
Histogram equalization  &  0.9995&1.0%k=5000
\\%k=16000(valeur optimale)
\hline

Gaussian noise ($\mu=0 $  $ ,\sigma=0.001$)   & 0.9998&1.0%k=5000
\\%k=16000(valeur optimale)
\hline
Gaussian noise ($\mu=0 $  $ ,\sigma=0.005$)  & 0.9928&0.9947%k=5000
\\ %k=16000(valeur optimale)
\hline
Salt \& pepper noise  ($\mu=0 $  $ ,\sigma=0.001$)  & 0.9997&1.0%k=5000
\\%k=16000(valeur optimale)
\hline
Low-pass Gaussian filtering  & & \\
 ($\sigma=0.5, 3 \times 3$)  & 0.9999 & 1.0 \\
 ($\sigma=0.5, 5 \times 5$)  & 0.9999 & 1.0\\
 ($\sigma=0.5, 7 \times 7$)  & 0.9998 & 1.0 \\
 ($\sigma=0.5, 9 \times 9$)  & 0.9998 &1.0 \\
 ($\sigma=0.6, 3 \times 3$)  & 0.9998 & 1.0 \\
 ($\sigma=0.6, 5 \times 5$)  & 0.9997 & 0.9997\\
($\sigma=0.6, 7 \times 7$) & 0.9996 & 0.9989 \\
($\sigma=0.6, 9 \times 9$)  & 0.9996&0.9984
%k=5000
\\%k=16000(valeur optimale)
\hline
Gaussian smoothing  &  & \\ 

  ($\sigma=0.5, 3 \times 3$)   & 1.0 &1.0 \\
 ($\sigma=0.5, 5 \times 5$)  & 0.9999  & 1.0 \\
 ($\sigma=0.5, 7 \times 7$) & 0.9999  & 0.9999\\
 ($\sigma=0.5, 9 \times 9$)  & 0.9999 &0.9999 \\ 
  ($\sigma=0.6, 3 \times 3$)   & 0.9996 &0.9954 \\
 ($\sigma=0.6, 5 \times 5$)  & 0.9996  & 0.9938 \\
 ($\sigma=0.6, 7 \times 7$)  & 0.9996  & 0.9883\\
 ($\sigma=0.6, 9 \times 9$)  & 0.9995 &0.9802%k=5000
\\ %k=16000(valeur optimale)
\hline
JPEG compression  & &  \\  
90\% & 0.9997 &1.0 \\
%85\% & 0.9995 &1.0 \\
80\% &0.9986 & 1.0 \\
75\% &0.9967 &1.0 \\
70\% & 0.9956&0.9959 \\
%65\% & 0.9940 &0.9943 \\
60\% & 0.9915 &0.9918 \\
%55\% & 0.9899 &0.9902 \\
50\% & 0.9871 &0.9874 \\
%45\% & 0.9705 &0.9708 \\
40\% & 0.9623&0.9626 \\
%35\% & 0.9564 &0.9567 \\
30\% & 0.9341 &0.9344 \\
%25\% & 0.8740 &0.8743 \\
20\% & 0.7887 &0.7890 \\
%15\% & 0.7482 &0.7485 \\
10\% & 0.6740 &0.6743 \\
\hline
JPEG2000 compression  & &  \\  
CR=2 & 0.9997 &1.0 \\
CR=4 & 0.9997 &1.0 \\
CR=6 & 0.9762 &0.9765 \\
CR=8 & 0.9460 &0.9463 \\
CR=10 &0.9002 &0.9003\\
\hline
Cropping & &\\
25\%& 1.0 &1.0 \\
50\% & 1.0 &1.0 \\
\hline
Combination attacks  & &\\  
HE + GN ($\sigma = 0.001$) & 0.9995&1.0 \\
HE + GN ($\sigma = 0.01$)& 0.9995&0.9882 \\
HE + SPN ($\sigma = 0.001$) &0.9995 &1.0 \\
HE + SPN ($\sigma = 0.01$)&0.9995 &0.9796 \\
GN ($\sigma = 0.001$) + JPEG compression (90\%) & 0.9995 &1.0 \\
GN ($\sigma = 0.001$) + JPEG compression (70\%) &  0.9994 & 0.9704 \\
SPN ($\sigma = 0.001$) + JPEG compression (90\%) &0.9997 &1.0 \\
SPN ($\sigma = 0.001$) + JPEG compression (70\%) & 0.9996 & 0.9946 \\
LPGF ($\sigma=0.5$, window size($9 \times 9$)) + SPN ($\sigma = 0.001$)  &0.9996 &0.9947 \\
LPGF ($\sigma=0.6$, window size($9 \times 9$)) + SPN ($\sigma = 0.001$)  & 0.9994 & 0.9694 \\
\noalign{\smallskip} \hline
\end{tabular}
\end{table}

\begin{table}[!]
%\footnotesize
%\centering
%{\renewcommand{\arraystretch}{1.6}
%{\setlength{\tabcolsep}{0.5cm} 
\caption{NC and PSNR values under various attacks for Mandrill}
\label{tab:RobustnessOfNonTexturedImagesMandrill}
\begin{tabular}{ccc}
\hline \noalign{\smallskip}
 & \multicolumn{2}{c}  {Proposed scheme}     \\
\cline{2-3}
  Attacks & \: NC \: \\
\hline
No Attack  & 1.0 \\  
\hline
  Histogram equalization  & 1.0  \\
\hline
  Gaussian noise  && \\
  ($\mu=0 $  $ ,\sigma=0.001$)  & 1.0 \\ 
($\mu=0 $  $ ,\sigma=0.005$)  & 1.0 \\ 
 ( $\mu=0 $  $ ,\sigma=0.01$)  & 1.0 \\ 
  ($\mu=0 $  $ ,\sigma=0.02$)  & 0.9857 \\ 
  ($\mu=0 $  $ ,\sigma=0.1$) & 0.9417 \\  
  \hline
  Salt \& Pepper & &  \\
  
  ($\mu=0 $  $ ,\sigma=0.001$)  & 1.0 \\
  ($\mu=0 $  $ ,\sigma=0.005$)  & 1.0 \\ 
  ($\mu=0 $  $ ,\sigma=0.01$)  & 0.9997 \\
  ($\mu=0 $  $ ,\sigma=0.02$) & 0.9843 \\ 
%  ($\mu=0 $  $ ,\sigma=0.03$) & 78.38 & 0.9798 \\
  \hline
  Low-pass Gaussian filtering  & \\
  ($\sigma=0.5$, $3 \times 3$) & 1.0   \\
  ($\sigma=0.5$, $5 \times 5$) & 1.0   \\
  ($\sigma=0.5$, $7 \times 7$) & 1.0   \\
  ($\sigma=0.5$, $9 \times 9$) & 0.9999  \\
  %\hline
 % Low-pass Gaussian filtering && \\
  ($\sigma=0.6$, $3 \times 3$) & 0.9834   \\
  ($\sigma=0.6$, $5 \times 5$) & 0.9818   \\
  ($\sigma=0.6$, $7 \times 7$) & 0.9794   \\
  ($\sigma=0.6$, $9 \times 9$) & 0.9765  \\
  \hline
 Gaussian smoothing  & \\
  ($\sigma=0.5$, $3 \times 3$)  & 1.0   \\
  ($\sigma=0.5$, $5 \times 5$)  & 1.0  \\
  ($\sigma=0.5$, $7 \times 7$) & 1.0   \\
  ($\sigma=0.5$, $9 \times 9$)  & 1.0   \\
  %  \hline
 % Gaussian smoothing  && \\
  ($\sigma=0.6$, $3 \times 3$)  & 0.9903  \\
  ($\sigma=0.6$, $5 \times 5$)  & 0.9838  \\
  ($\sigma=0.6$, $7 \times 7$)  & 0.9784   \\
  ($\sigma=0.6$, $9 \times 9$)  & 0.9741   \\
    \hline
  
  Cropping  &  \\ 

  ($10\%$)   &1.0   \\
  ($25\%$) & 1.0     \\
  ($50\%$) & 0.9999     \\ 
  \hline 
  Combined attacks & \\

\: HE + GN ($\mu=0 $ , $\sigma=0.001$) \:  & 1.0 \\ %k=9000

 \: HE + GN ($\mu=0 $ , $\sigma=0.01$) \:  & 0.9875 \\ %k=9000

  HE + SPN ($\sigma=0.001$)  & 1.0   \\
  
 HE + SPN ($\sigma=0.01$) &  0.9886   \\  
  
 GN ($\mu=0 $  $ ,\sigma=0.001$) + JPEG compression (QF=90)  & 1.0  \\

 GN ($\mu=0 $  $ ,\sigma=0.001$) + JPEG compression (QF=70)  & 0.9715  \\
  
  SPN($\sigma=0.001$) + JPEG compression (QF=90)  & 1.0   \\

 SPN($\sigma=0.001$) + JPEG compression (QF=70)  & 0.9892   \\

LPGF( $\sigma= 0.5$, window size ($9 \times 9$) ) + SPN ($\sigma=0.01$) & 0.9999 \\

LPGF ( $\sigma= 0.6$, window size ($9 \times 9$) ) + SPN($\sigma=0.01$) & 0.9703 \\

%Low-pass Gaussian filtering ( $\sigma= 0.6$, window size ($9 \times 9$) ) + Salt \& pepper noise ($\sigma=0.001$)& 79.99 & 0.9999 \\

\noalign{\smallskip} \hline
\end{tabular}
\end{table}

\begin{table}[!]
%\footnotesize
\centering
%{\renewcommand{\arraystretch}{1.9}
%{\setlength{\tabcolsep}{0.7cm} 
\caption{Robustness comparison between the DFT only and the  DFT-DCT  based algorithm for  Mandrill in terms of  NC}
\label{tab:RobustnesscomparisonforDFTonlyAndDftDct}
\begin{tabular}{ccc}
\hline \noalign{\smallskip}
% & \multicolumn{2}{c|}  {Images}     \tabularnewline
%\cline{2-3}
 & \multicolumn{2}{c}  {Watermarking methods}     \\
\cline{2-3}
  \: Attacks
   \: &  DFT  only  & DFT-DCT\\ 
\hline

 & \multicolumn{2}{c}  {Robustness metric}     \\
\cline{2-3}
   & NC & NC \\ 
\hline
No attack&1.0  & 1.0 \\
GN $(\mu=0, \sigma =0.001)$&0.7140 &1.0 \\
GN $(\mu=0, \sigma =0.005)$&0.6872 &1.0 \\
SPN $(\mu=0, \sigma =0.001)$&0.7365 & 0.9947 \\
JPEG $(90\%)$  & 0.6854  &1.0 \\
%JPEG $(85\%)$  & 0.6780 &1.0 \\
JPEG $(80\%)$  & 0.6476  &1.0 \\
JPEG $(75\%)$  & 0.6324  &1.0 \\
JPEG $(70\%)$ & 0.6003  &0.9998 \\
%JPEG $(65\%)$  &0.5488  &0.9953 \\
JPEG $(60\%)$  & 0.5243  &0.9843 \\
%JPEG $(55\%)$  & 0.5177  &0.9808 \\
JPEG $(50\%)$  & 0.5102  &0.9788 \\
%JPEG $(45\%)$  & 0.5062  &0.9646 \\
JPEG $(40\%)$  & 0.5035  &0.9584 \\
%JPEG $(35\%)$  & 0.4843  &0.9402 \\
JPEG $(30\%)$  & 0.4690  &0.9261 \\
%JPEG $(25\%)$  & 0.4473  &0.7875 \\
JPEG $(20\%)$  & 0.4198 &0.7795 \\
%JPEG $(15\%)$  & 0.3845  &0.7308 \\
JPEG $(10\%)$  & 0.3268  &0.6433 \\
JPEG2000 (CR$=2)$  & 0.6876  &1.0 \\
JPEG2000 (CR$=4$)  & 0.6134  &1.0 \\
JPEG2000 (CR$=6$) & 0.6068  &0.9873 \\
JPEG2000 (CR$=8$) & 0.5846  &0.9532 \\
JPEG2000 (CR$=10$)  & 0.5243  &0.9118 \\
LPGF $(\sigma=0.5, 3\times3)$ & 0.7865  &1.0 \\
LPGF $(\sigma=0.5, 5\times5)$&  0.7369 &1.0 \\
LPGF $(\sigma=0.5, 7\times7)$ & 0.7166  &1.0 \\
LPGF $(\sigma=0.5, 9\times9)$&  0.7087 &0.9999 \\ 
GS $(\sigma=0.5, 3\times3)$ & 0.7750  & 1.0 \\
GS $(\sigma=0.5, 5\times5)$ & 0.7434  & 1.0 \\ 
GS $(\sigma=0.5, 7\times7)$ & 0.7190  & 1.0 \\ 
GS $(\sigma=0.5, 9\times9)$ & 0.7003  & 1.0 \\   
HE  &0.9720&1.0 \\
Cropp (10\%)& 0.9986   & 1.0 \\
Cropp (20\%)& 0.9932   &1.0 \\
Cropp (25\%)& 0.9890   & 1.0 \\
Cropp (40\%)& 0.9878   & 1.0 \\
Cropp (50\%)& 0.9689   & 0.9999 \\
Rotation $(\theta = 0.25\degree)$ & 1.0  & 1.0 \\
Rotation $(\theta = 0.75\degree)$ & 1.0  & 0.9999 \\
Rotation $(\theta = -0.25\degree)$& 1.0  & 1.0 \\
Rotation $(\theta = -0.75\degree)$& 1.0  & 0.9998 \\
HE+GN $(\sigma =0.001)$& 0.7387 &1.0 \\
HE+SPN $(\sigma =0.001) $& 0.7407  &1.0 \\  
GN $(\sigma =0.001)$+JPEG $(90\%) $ & 0.6084 & 1.0 \\
SPN $(\sigma =0.001)$+JPEG$(90\%)$  & 0.5435& 1.0 \\
LPGF $(9\times9)$+SPN$(\sigma =0.001)$  & 0.5108&0.9999 \\
Average   & 0.7331 &0.9795
\\
\noalign{\smallskip} \hline
\end{tabular}
\end{table}
In order to evaluate the robustness of the proposed  scheme, we calculate the normalized correlation (NC) between the original watermark and the extracted one. %and the PSNR between the extracted watermark and the original watermark, respectively.
Before applying  attacks, it can be observed that the watermark was extracted perfectly with a correlation NC$=1$.
%Moreover, the PSNR value between the original watermark and the extracted watermark is $80$ dB  which reflects a good extraction of the watermark in the absence of attacks.
To test the algorithm robustness, the watermarked images are exposed to various attacks: 1) noising attack : Gaussian Noise (GN) and salt \& pepper noise (SPN); 2) format-compression attack : JPEG and JPEG2000 compression; 3) image-processing attack : low-pass Gaussian Filtering (LPGF), Gaussian smoothing(GS), and histogram equalization (HE);  4) Geometric distortion: cropping (Cropp) and rotation.
5) Combined attacks : histogram equalization \& Gaussian noise (HE+GN), histogram equalization and salt \& pepper noise (HE+SPN), Gaussian noise \& JPEG (GN+JPEG), salt \& pepper noise and JPEG (SPN+JPEG), low-pass Gaussian filtering and salt \& pepper noise (LPGF+SPN).
Fig. \ref{fig:Robustness_in_terms_of_NC_after_several_attacks_applied_to_a_simple_of_textured_images} represents the obtained NC values after a wide rang of attacks applied to a simple of textured images taken from Brodatz database \cite{RefJ31}. 
It can be seen from Table \ref{tab:RobustnesscomparisonforDFTonlyAndDftDct},  Table \ref{tab:SSIM_and_NC_value_after_several_attacks_TexturedImg.}
 and Fig. \ref{fig:Robustness_in_terms_of_NC_after_several_attacks_applied_to_a_simple_of_textured_images} that the proposed scheme is very robust to histogram equalization regardless of the image nature.
Fig. \ref{fig-ExtractedWatermarksAfterSeveralAttacks} displays the extracted watermarks after several attacks (Histogram equalization , Salt \& Pepper noise, JPEG compression, Gaussian smoothing, cropping,etc.). We can see visually that although the watermarked images are exposed to these attacks, the watermark is almost extracted perfectly.

%As seen, in all attacks, the proposed scheme reaches high NC values getting more robustness against several attacks.
% Furthermore, all the obtained NC values are greater than $0.9742$ ( NC obtained after Cropping attack ($50\%$)).

%\textcolor{blue}{
In order to improve further the robustness of our approach, we compare the performance in terms of NC between the DFT-DCT and the DFT-only based method after carrying out several kind of attacks.  
It can be observed from Table \ref{tab:RobustnesscomparisonforDFTonlyAndDftDct} that the DFT-DCT approach improved the robustness performance considerably when compared to the DFT-only watermarking method.
%J'ai enlev?? cette figure car dans ses remarques ce graphe n'a pas de signiification car on a la meme valeur de 1.
%\begin{figure}
%\centering
%\includegraphics[scale=0.55]{figs/Robustness_in_terms_of_NC_after_gaussian_smoothing_for_non_textured_images.eps}
%\caption{Robustness in terms of NC after Gaussian smoothing for natural images.}
%\label{fig:Robustness in terms of NC after Gaussian smoothing for natural images.}
%\end{figure}

\subsubsection*{5.2.2.1 Noising attack}
 First, we carried out the addition of Gaussian noise with zero mean ($\mu=0$) and several variance values (GN1 : $\sigma=0.001$, GN2 : $\sigma=0.005$, $\sigma=0.01$, $\sigma=0.02$ and $\sigma=0.1$) in order to better understand the limitations of the proposed method. As reported in Fig. \ref{fig:Robustness_in_terms_of_NC_after_several_attacks_applied_to_a_simple_of_textured_images} and Table \ref{tab:RobustnessOfNonTexturedImagesMandrill}, it appears that for  variance values below $0.02$ the proposed scheme is quite robust to noise addition (NC$=1.0$). For higher variance values, the NC values decreases slightly but the results are still good ($\sigma=0.1$, NC$=0.9417$). 
 % ( NC$=0.9857$ for $\sigma=0.02$ and NC$=0.9417$ for $\sigma=0.1$ ). \\
  Second, Salt \& Pepper  noise (SPN) has also been applied with zero mean ($\mu=0$) and  several intensities ($\sigma=0.001$, $\sigma=0.005$, $\sigma=0.01$ and $\sigma=0.02$) with the aim of analyzing the limitation of the proposed work. It clearly appears from Fig. \ref{fig:Robustness_in_terms_of_NC_after_several_attacks_applied_to_a_simple_of_textured_images}   and Table \ref{tab:RobustnessOfNonTexturedImagesMandrill} that the method show good robustness against salt \& pepper noise for intensity values less than  $0.02$. For an intensity value equal to $0.02$ the results are encouraging (NC$=0.9843$).
\subsubsection*{5.2.2.2 Compression attack} 
\begin{figure}[!h]
\centering
\subfigure[]{\label{sub100}\includegraphics[width=0.65\textwidth]{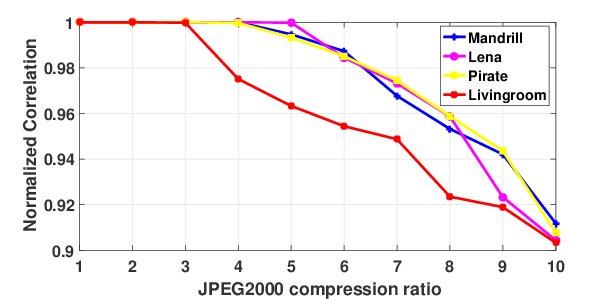}}
\subfigure[]{\label{sub101}\includegraphics[width=0.65\textwidth]{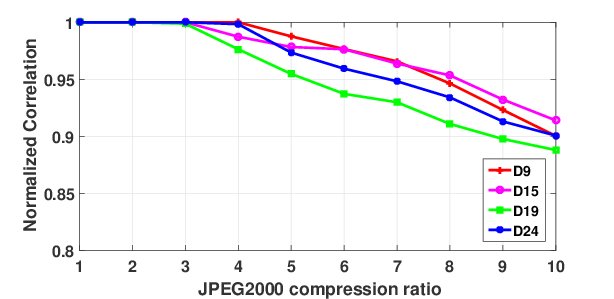}}
\caption{(a) : robustness in terms of NC after JPEG2000 compression attack applied to a simple of natural images, (b): robustness in terms of NC after JPEG2000 compression attack applied to a simple of textured images }
\label{fig:NCRobustJPEG2000NaturTextImages}
\end{figure}

\begin{figure}[!]
    \centering
    \subfigure[]{\label{sub1} \includegraphics[width=2.4cm]{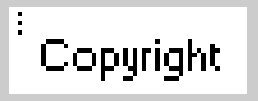}}
    \subfigure[]{\label{sub2} \includegraphics[width=2.4cm]{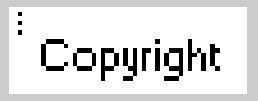}}
    \subfigure[]{\label{sub3} \includegraphics[width=2.4cm]{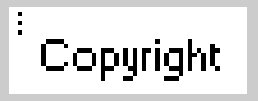}}
    \subfigure[]{\label{sub4} \includegraphics[width=2.4cm]{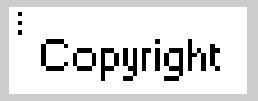}}
    \subfigure[]{\label{sub5} \includegraphics[width=2.4cm]{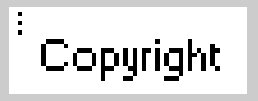}}
    \subfigure[]{\label{sub6} \includegraphics[width=2.4cm]{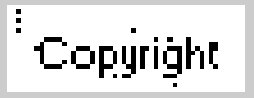}}
    \subfigure[]{\label{sub7} \includegraphics[width=2.4cm]{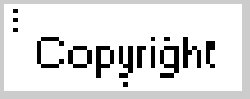}}
    \subfigure[]{\label{sub8} \includegraphics[width=2.4cm]{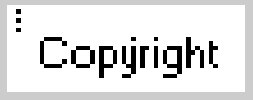}}
\subfigure[]{\label{sub9} \includegraphics[width=2.4cm]
{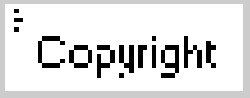}}
\subfigure[]{\label{sub10} \includegraphics[width=2.4cm]{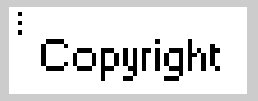}}
\subfigure[]{\label{sub11} \includegraphics[width=2.4cm]{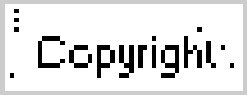}}
\subfigure[]{\label{sub12} \includegraphics[width=2.4cm]{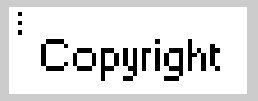}}
\subfigure[]{\label{sub13} \includegraphics[width=2.4cm]{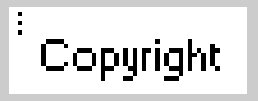}}
\subfigure[]{\label{sub14} \includegraphics[width=2.4cm]{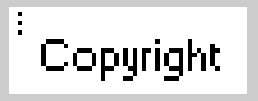}}
\subfigure[]{\label{sub15} \includegraphics[width=2.4cm]{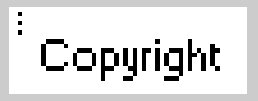}}
\subfigure[]{\label{sub16} \includegraphics[width=2.4cm]{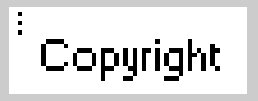}}

    \caption{Extracted watermarks after attacks : (a) Embedded Watermark,(b) HE, (c) GN($\sigma=0.001$), (d) SPN , (e) LPGF ( window size : ($9 \times 9$)), (f) JPEG (Q$=60$) , (g) JPEG (Q$=65$)  , (h) JPEG (Q$=70$),  (i) JPEG (Q$=75$), (j) Gaussian Smoothing (window size : ($9 \times 9$)), (k) Cropping $50\%$, (l) Cropping $25\%$, (m) HE+ SPN ($\sigma = 0.001$), (n) GN ($\sigma=0.001$) + JPEG (QF$=90$), (o) SPN ($\sigma =0.001$) + JPEG (QF$=90$), (p) HE + GN ($\sigma=0.001$) }
    \label{fig-ExtractedWatermarksAfterSeveralAttacks}
\end{figure}

Robustness against lossy compression is of crucial importance due to the wide diffusion of lossy compression tools and the huge use of this image format.
 To assess the performance from this point of view, we iteratively applied JPEG compression to the watermarked images, each time increasing the quality factor, i.e decreasing the compression ratio, ranging from $10$ to $90$. 
 Note that the quality factor for images is an integer value ranging from $1$ to $99$, which denotes the predetermined image quality.
 Moreover, robustness against JPEG2000  using different compression ratios has been investigated. The compression ratio (CR) is varied from $1$ to $10$. Fig. \ref{fig:NCRobustJPEG2000NaturTextImages} shows the robustness in terms of NC against JPEG2000 compression for both natural and textured images. It can be seen from Fig. \ref{fig:NCRobustJPEG2000NaturTextImages}, Tables \ref{tab:SSIM_and_NC_value_after_several_attacks_TexturedImg.} and \ref{tab:RobustnesscomparisonforDFTonlyAndDftDct} that the NC values are close to $1$ when CR$<5$. By increasing the compression ratio, the NC values decreases slightly but the results are still encouraging (NC$=0.9$ when CR$=10$). 
  
%\begin{table}[H]
%\footnotesize
%\centering
%%{\renewcommand{\arraystretch}{1.2} 
%\caption{NC values after JPEG 2000 with several compression ratios 5,10,20,30,40,50,60,70,80 in Lena, Mandrill, pirate and Walkbridge.}
%\label{tab:jpeg2000Robustness}
%%R??sultats pour l'image baboon
%\begin{tabular}{ccccc}
%\hline \noalign{\smallskip}
%\: JPEG 2000 Compression \: & \:  Lena  \: & \: Mandrill \:& Pirate & Walkbridge \\
% \hline
%CR=$5$ & 1.0 &1.0&1.0&1.0\\
%CR=$10$ & 1.0 & 1.0 &1.0& 1.0\\
%CR=$20$ & 1.0 & 0.97 &1.0&1.0\\
%CR=$30$ & 0.9980 & 0.86&1.0&1.0\\
%CR=$40$ & 0.9384 & 0.79 &1.0&1.0\\
%CR=$50$ & 0.7382 & 0.66 &1.0&1.0\\
%CR=$60$ & 0.6545 & 0.61 &1.0&1.0\\
%CR=$70$ & 0.7382 & 0.66 &1.0&1.0\\
%CR=$80$ & 0.6545 & 0.61 &1.0&1.0\\
%\end{tabular}
%\end{table}

%JPEG2000 is based on wavelet coding
%techniques and extends the initial JPEG standard
%Some of the advantages of JPEG2000 standard are
%its superior low bit-rate performance and its robustness to bit-errors.
 
  Afterwards, we compare the proposed method to 
%\cite{lien2006watermarking},
 \cite{RefJ14}, \cite{RefJ15}, \cite{RefJ16} , \cite{RDWTMTAP2017}, \cite{SinghAK2015}   and \cite{SinghAK2016}.
 %(see Table  \ref{tab:Robustnesscomparisonforlena}).

%In order to evaluate the robustness against JPEG compression, we compressed the watermarked images "Mandrill", "Peppers", "cameraman", "Lena" and "Goldhill" by different quality factors and we summarize the average NC values for the five test images. 
Fig. \ref{fig:Robustness in terms of NC after JPEG compression attack applied to a simple of textured image.} and Fig. \ref{fig:Robustness in terms of NC after JPEG compression attack applied to a simple of natural image.} summarized the results obtained in terms of  NC after JPEG compression using several quality factors for a sample of textured images taken from Brodatz \cite{RefJ31} and five natural images, respectively. 
The first observation to make after looking at the results is that there is a small difference between the NC values, from natural images to textured images even if each kind of image has its specific characteristics.  As depicted in Table  \ref{tab:RobustnesscomparisonforDFTonlyAndDftDct} and Table \ref{tab:SSIM_and_NC_value_after_several_attacks_TexturedImg.}, the NC value is equal to $1$ when the quality factor greater than $75$. Otherwise, the NC value  decreases but the results are encouraging and  proved that the proposed technique is still robust to JPEG compression (quality factor$=30$, NC$=0.9344$ ) for D9 image, and (quality factor$=30$, NC$=0.9261$ ) for Mandrill image.
As shown in Table \ref{tab:RobustnesscomparisonforDFTonlyAndDftDct}
the results obtained in terms of robustness against JPEG of the DCT-DFT method are good and outperform the DFT-only method. 
%We believe that the robustness against JPEG compression is achieved by applying the DCT to the DFT magnitude.

%\textcolor{blue}{
% Furthermore, Table \ref{tab:SSIM_and_NC_value_after_several_attacks_TexturedImg.} show the  NC values, which are almost close to $1$, obtained after JPEG attack with their corresponding PSNR values in the case of "D9".}
 
%  It can be seen from Table \ref{tab:jpegcompressionattackcomparaison2} that our method gives better results than the approach in \cite{sahraee2013robust}. In addition, as depicted in Table \ref{tab:Robustnesscomparisonforlena}, it can be observed that our scheme outperforms schemes 
% %\cite{lien2006watermarking},
%  \cite{wang2004wavelet} and \cite{sahraee2013robust}.
%phrases ?? ajouter dans cette partie:
% the watermark could still be well detected even
%after the image was compressed using a compression
%ratio of 70 
%Et enchainer : Also the proposed scheme was found
%to be robust to low pass filtering as shown in Fig. ...

% \par
%ajouter cette phrase 'we note that in the last two combined attacks the quality factor of JPEG is set to 90%

\begin{figure}
    \centering
    \subfigure[]{\label{sub31} \includegraphics[width=0.17\textwidth]{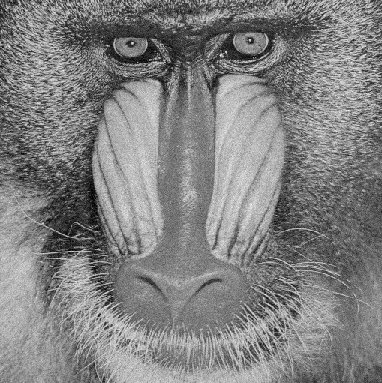}}
    \subfigure[]{\label{sub32} \includegraphics[width=0.17\textwidth]{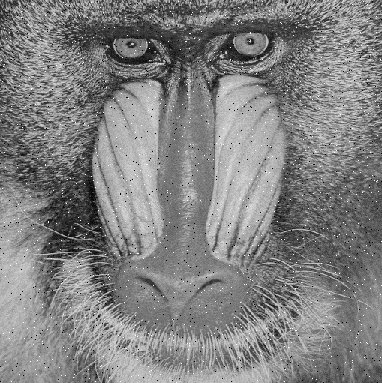}}
        \subfigure[]{\label{sub33} \includegraphics[width=0.17\textwidth]{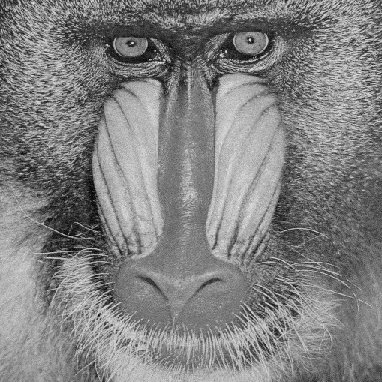}}
        \subfigure[]{\label{sub34} \includegraphics[width=0.17\textwidth]{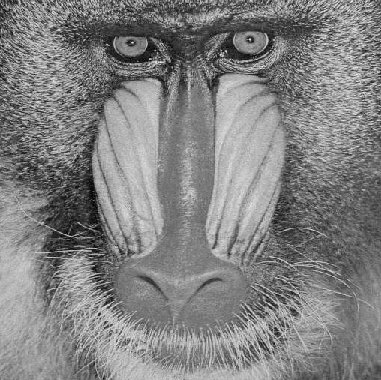}}
        \subfigure[]{\label{sub35} \includegraphics[width=0.17\textwidth]{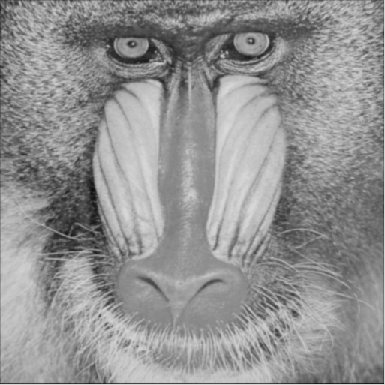}}
        \subfigure[]{\label{sub36} \includegraphics[width=0.17\textwidth]{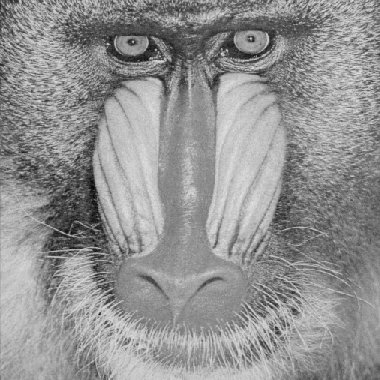}}
        \subfigure[]{\label{sub37} \includegraphics[width=0.17\textwidth]{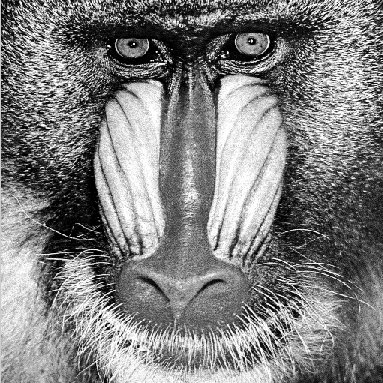}}
        \subfigure[]{\label{sub38} \includegraphics[width=0.17\textwidth]{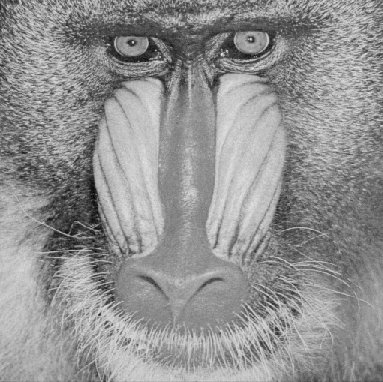}}       
        \subfigure[]{\label{sub39} \includegraphics[width=0.17\textwidth]{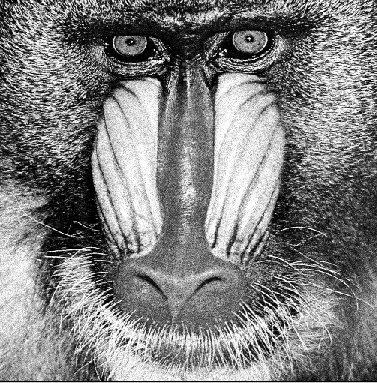}}
        \subfigure[]{\label{sub40} \includegraphics[width=0.17\textwidth]{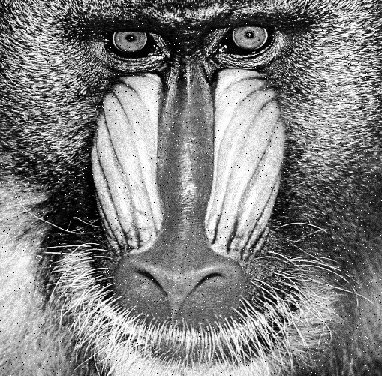}}
        \subfigure[]{\label{sub41} \includegraphics[width=0.17\textwidth]{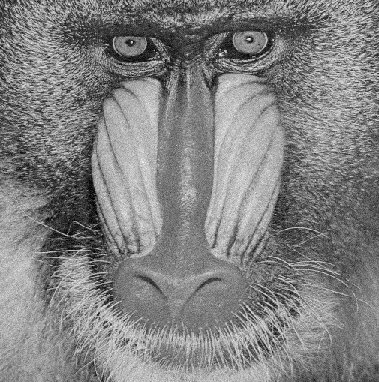}}
        \subfigure[]{\label{sub42} \includegraphics[width=0.17\textwidth]{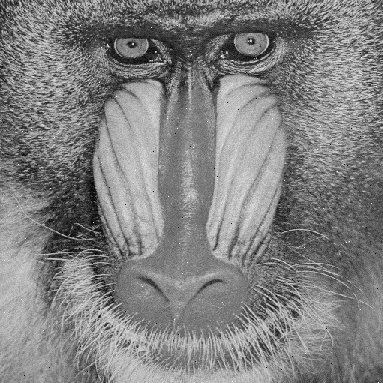}}
\subfigure[]{\label{sub42} \includegraphics[width=0.17\textwidth]{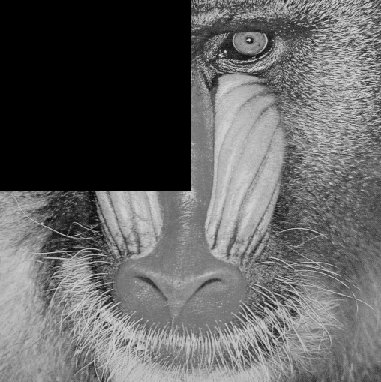}}
\subfigure[]{\label{sub43} \includegraphics[width=0.17\textwidth]{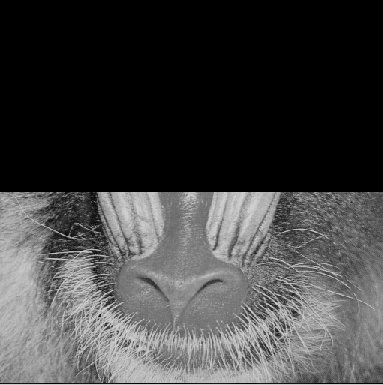}}
       \subfigure[]{\label{sub44}        
        \includegraphics[width=0.17\textwidth]{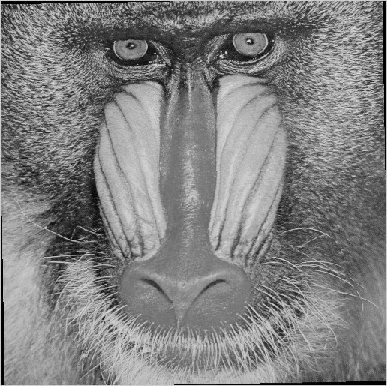}}         
    \caption{Sample of attacked watermarked images : (a) Gaussian noise addition with mean zero and standard deviation 0.001, (b) Salt \& pepper noise with noise density $0.01$, (c) JPEG compression with quality factor $90$ , (d) JPEG compression with quality factor $60$, (e) Gaussian low passe filtering with window size $3 \times 3$, (f) Gaussian low passe filtering with window size $9 \times 9$, (g) Histogram equalization, (h) Gaussian smoothing with window size $9 \times 9$, (i) Histogram equalization + Gaussian noise addition with mean zero and standard deviation $0.001$, (j) Histogram equalization + Salt \& pepper noise with noise density $0.001$, (k) Gaussian noise addition with mean zero and standard deviation $0.001$ + JPEG compression with quality factor $90$, (l) Salt \& pepper noise with noise density $0.01$ + JPEG compression with quality factor $90$, (m) Cropping $25\%$, (n) Cropping $50\%$,  (o) Rotation $1\degree$ }
    \label{fig-AttackedWatermarkedImages}
\end{figure}
%\begin{table}[!t]
%\footnotesize
%\centering
%{\renewcommand{\arraystretch}{1.2} %donne la distance entre les lignes%
%%{\setlength{\tabcolsep}{0.6cm} %donne la distance entre les collones%
%\caption{PSNR and NC values after JPEG Compression attack : (QF=90), (QF=85), (QF=80), (QF=75), (QF=70), (QF=65) and (QF=60).}
%\label{tab:jpegcompressionattack}
%%R??sultats pour l'image baboon
%\begin{tabular}{|c||c||c|}
%\hline
%\: JPEG Compression \: & \:  PSNR(dB) \: & \: NC \: \tabularnewline
% \hline
%(QF=90) & 35.63 & \textbf{1.0}%k=5000
%\tabularnewline%k=16000(valeur optimale)
%\hline
%(QF=85) & 33.20 & \textbf{1.0} \tabularnewline% valeur optimale est k=6000
%\hline
%(QF=80)& 31.41 & \textbf{1.0}\tabularnewline
%\hline
%(QF=75) & 30.31 & \textbf{1.0}\tabularnewline%k=11000
%\hline%pour k=8000, psnr=30.60, NC=0.9895
%(QF=70) & 29.87 & \: \textbf{0.9840} \: \tabularnewline
%\hline% valeur optimale est k=8500
%(QF=65) & 29.14 & \: \textbf{0.9738} \: \tabularnewline
%\hline%k=14000
%(QF=60) & 28.46 & \: \textbf{0.9343} \: \tabularnewline
%\hline %k=18000
%\end{tabular}}
%\end{table}

\subsubsection*{5.2.2.3 Low-pass Gaussian filtering}
The low-pass Gaussian filtering attack is also one of the common manipulations in image processing. It aims to remove high frequency components from the image. 
The watermarked images ware filtered with a low-pass Gaussian filter using several window sizes $(3\times3)$, $(5\times5)$, $(7\times7)$ and $(9 \times 9)$ and two standard deviation values ($\sigma=0.5, \sigma=0.6$).
 The results shown in Table \ref{tab:RobustnessOfNonTexturedImagesMandrill} and Table \ref{tab:SSIM_and_NC_value_after_several_attacks_TexturedImg.} in    terms of NC and PSNR are obtained after applying the low-pass Gaussian filtering to Mandrill natural image and D9 textured image taken from Brodatz \cite{RefJ31}.
 From Table \ref{tab:RobustnessOfNonTexturedImagesMandrill} and Table \ref{tab:SSIM_and_NC_value_after_several_attacks_TexturedImg.} , it is clear that our approach is  robust to low-pass Gaussian filtering. The results show that the robustness is still good even with larger filter size. As depicted in Table \ref{tab:RobustnessOfNonTexturedImagesMandrill}, in the case of Mandrill image, with a standard deviation $\sigma =0.6$ and filter size $(9\times9)$, the obtained NC and PSNR are $0.9765$ and $78.12$ dB, respectively. Similarly, in the case of D9 textured image, it can be observed from Table  \ref{tab:SSIM_and_NC_value_after_several_attacks_TexturedImg.}  that with a standard deviation $\sigma =0.6$ and filter size $(9\times9)$, the obtained normalized correlation  (NC) is $0.9984$.    

\subsubsection*{5.2.2.4  Gaussian smoothing}
Gaussian smoothing is a very common operation in image processing. It consists of removing detail and noise.
We have applied the Gaussian smoothing attack to the images test with different standard deviations and window sizes. From Table \ref{tab:RobustnessOfNonTexturedImagesMandrill} and Table \ref{tab:SSIM_and_NC_value_after_several_attacks_TexturedImg.} it can be seen that the proposed method is  robust against Gaussian smoothing attack for several filter sizes ($(3 \times 3)$, $(5 \times 5)$, $(7 \times 7)$ and $(9 \times 9)$). The results obtained in terms of NC  are close to $1$. In fact, even with  ($\sigma=0.6$) and size window $(9\times9)$ , NC=$0.9741$. In addition, as depicted in Table \ref{tab:RobustnesscomparisonforDFTonlyAndDftDct}, the proposed system not only shows good resistance against Gaussian smoothing but outperforms the DFT-only approach. 
\subsubsection*{5.2.2.5 Cropping}
%It can be seen from Fig . \ref{fig:Robustness in terms of NC after cropping attack applied to a simple of natural image.} that even the test images are 50\% cropped, the watermark can still be identified easily. 
%According to Table \ref{tab:Robustnesscomparisonforlena}, the proposed scheme, compared with the other schemes, gives the best performance. 

Image cropping is one of the  most common manipulations in digital image. It's the most severe geometric distortion to be applied against an image. It consists of cropping off a rectangular region of the image by setting its pixels to zero value.
 To check the robustness of our proposal, we apply cropping attacks with several proportions  ($10\%$, $20\%$, $25\%$, $40\%$ and $50\%$ ) to the watermarked images then watermark is extracted.
It can be concluded from Fig. \ref{fig:Robustness in terms of NC after cropping attack applied to a simple of natural image.} that the proposed scheme is very robust to the cropping attack.
Table \ref{tab:RobustnessOfNonTexturedImagesMandrill} and Table \ref{tab:SSIM_and_NC_value_after_several_attacks_TexturedImg.} show the obtained results after applying cropping to "Mandrill" and "D9" respectively. The above results in terms of NC, under cropping $50\%$, show that our method is able to withstand this attack(NC=$1.0$).   
 The main reason stands on the fact that the effect of cropping leads to the blurring of spectrum. Therefore,  there is no need of any synchronization since the watermark is embedded in the DFT  magnitude.
%  the magnitude of the Fourier transform is invariant to translation in the spatial domain which leads automatically to the fact that the watermark resist to cropping attack. }

% Finally, from table \ref{tab:tableau-comparatif}, it can be observed that our approach is also robust against combined attack ( Histogram equalization + Gaussian noise ) with (NC=1).
\subsubsection*{5.2.2.6 Combined attacks} 
 The goal of this experiment is to check whether this kind of combination attack is able to remove the watermark of the proposed scheme. To test further the robustness of our method, different combinations of attacks composed by several kinds of attacks have been carried out. Table \ref{tab:SSIM_and_NC_value_after_several_attacks_TexturedImg.}  and Table \ref{tab:RobustnessOfNonTexturedImagesMandrill}  sketch the NC values where we can see the robustness of our method both for textured and natural images. Moreover, it can be seen from Table \ref{tab:Robustnesscomparisonforlena} that the results obtained after rotation attack are encouraging.
\par 
 
In conclusion, it can be observed from Table \ref{tab:SSIM_and_NC_value_after_several_attacks_TexturedImg.}  and Table \ref{tab:RobustnessOfNonTexturedImagesMandrill} that, in all the cases, our method achieves good watermark extraction capability against several kind of attacks independently of the image nature.
That is illustrated by the obtained values of the  NC calculated between the original watermark and the extracted one which are above $0.9694$. 
Moreover, regardless of the attack type, it can be concluded that the obtained results  in terms of NC of the DFT-DCT method outperforms the performances of the DFT-only method.
\subsection{Computational complexity}
To evaluate the computational complexity of the proposed method, the complexity of $Algorithm 1$(watermark embedding) and $Algorithm 2$(watermark extraction) has been calculated  using the big $O$ notation. Due to the matrix multiplication involved in equation \ref{eq:Imagereconstruction}, the computational complexity of the proposed method is $O(n^3)$. In the extraction process, the  complexity is $O(n^2)$ due to inverse Arnold transform calculation. Consequently, the global complexity of the proposed method is $O(n^3)$.\\
In order to analyze the time complexity of the proposed scheme, several experiments have been conducted on $10$ natural images of size $512\times512$. Table \ref{tab:ComputationalTime} shows the average CPU time of the watermark embedding and extraction. The experiments are performed using MATLAB R2013a environment on a PC with CPU Intel(R) Core(TM) i5-3470 @ 3.2 GHZ with 4-GB of RAM.
%It can be observed that the computational time of the proposed scheme is reasonable
\begin{table}[H]
\footnotesize
\centering
%{\renewcommand{\arraystretch}{1.2}
\caption{Average CPU time for the proposed watermarking method}
\label{tab:ComputationalTime}
%R??sultats pour k=750
\begin{tabular}{lc}
\hline \noalign{\smallskip}
  Computational time (seconds)  &  The proposed method \\
 \hline
% \cite{lien2006watermarking} &  41.54  \tabularnewline
Embedding time & 0.9483  \\
Extraction time & 0.8845\\
Total time & 1.8328 \\
\noalign{\smallskip} \hline
\end{tabular}
\end{table} 

\subsection{Comparison with alternative methods}
To further demonstrate the robustness of the proposed method, we compare it with schemes \cite{RefJ14}, \cite{RefJ15}, \cite{RefJ16}, \cite{RDWTMTAP2017}, \cite{SinghAK2015} and \cite{SinghAK2016} in terms of imperceptibility and robustness as well as capacity.
\subsubsection{Imperceptibility}

In Table \ref{tab:imperceptibility2}, is presented the comparison in terms of imperceptibility between the proposed scheme and the schemes in 
%\cite{lien2006watermarking}, 
\cite{RefJ14}, \cite{RefJ15}, and  \cite{RefJ16}.  The metric used in comparison is PSNR using Lena as test image.
%Table 2 demonstrates the imperceptibility comparaison results of the proposed method and scheme [14].
 The  PSNR values show the superiority of our method even if its capacity is bigger than the alternative methods. 
 We believe that the main reason stands on the fact that the watermark is inserted in the DFT magnitude which ensures high imperceptibility.
%  \ref{tab:ImpercepcompaMtap2017Capacity256by256} a 
%  comparison in terms of PSNR and SSIM with scheme in \cite{RDWTMTAP2017} using a watermark of size $\256 times 256$. 
% 
\begin{table}[H]
\footnotesize
\centering
%{\renewcommand{\arraystretch}{1.2}
\caption{Watermark imperceptibility and capacity using Lena as test image}
\label{tab:imperceptibility2}
%R??sultats pour k=750
\begin{tabular}{ccc}
\hline \noalign{\smallskip}
 \: Watermarking methods  \: & \: PSNR (dB) \: & \: Capacity (bits)\:\\
 \hline
% \cite{lien2006watermarking} &  41.54  \tabularnewline
\cite{RefJ14}& 38.20  & 512\\
\cite{RefJ15} & 51.80 & 512\\
\cite{RefJ16} & 44.73 & 512\\
 \cite{RDWTMTAP2017} & 41.36&16834\\
 Proposed method & 61.97 & 988\\
\noalign{\smallskip} \hline
\end{tabular}
\end{table} 

\begin{table}[H]
%\footnotesize
\centering
%{\renewcommand{\arraystretch}{1.9}
%{\setlength{\tabcolsep}{0.7cm} 
\caption{Comparison of the imperceptibility in terms of PSNR and SSIM between the proposed method and \cite{RDWTMTAP2017}for several images using a watermark of size $256\times256$}
\label{tab:ImpercepcompaMtap2017Capacity256by256}
\begin{tabular}{ccccc}
\hline \noalign{\smallskip} 
% & \multicolumn{2}{c|}  {Images}     \tabularnewline
%\cline{2-3}
 & \multicolumn{4}{c}  {Watermarking methods}     \\
\cline{2-5}
  \: Cover image \: &  Scheme in \cite{RDWTMTAP2017}  &  & Proposed method  &\\ 
\hline

 & \multicolumn{4}{c}  {Imperceptibility metric}   \\
\cline{2-5}
  & \: PSNR \:  & SSIM &PSNR& SSIM\\
\hline
%No Attack  & Infinity & \textbf{1.0} &&&&\tabularnewline  
%\hline
Lena &39.774& 0.9937&40.93& 0.9972 \\
 Cameraman & 39.619 & 0.9868&42.03&0.9987\\
Mandrill  & 32.492 & 0.9611 &39.85&0.9803\\
Peppers & 39.031  &0.9831&43.21&0.9990\\
\noalign{\smallskip}\hline
\end{tabular}
\end{table}

  According to table \ref{tab:ImpercepcompaMtap2017Capacity256by256}, it can be seen that the proposed method gives good results in terms of imperceptibility with a capacity of $65536$ bits and outperforms the Singh \textit{et al.} \cite{RDWTMTAP2017}.
\subsubsection{Robustness}
~~\\

The robustness comparison is performed in the case of Gaussian noise, salt \& pepper noise, histogram equalization, JPEG compression, Cropping and rotation attacks.
%JPEG attack
%SPN attack
~~\\

For  Salt \& pepper attack, as  shown in Table \ref{tab:SPNComparison}, 
it can also be observed that, compared to scheme  \cite{RefJ15}, our approach is more robust. Moreover, it can be seen from Table \ref{tab:RobustnesscomparisonWithMTAP2017forlena} and Table  \ref{tab:RobustnesscomparisonMTAPmethodsforlena} that the proposed method shows high robustness to Salt \& pepper noise compared to \cite{RDWTMTAP2017}, \cite{SinghAK2015} and \cite{SinghAK2016}.

Moreover, the proposed technique is robust to histogram equalization and outperforms the schemes in \cite{RefJ15}, \cite{RefJ16}, \cite{RDWTMTAP2017}, \cite{SinghAK2015} and \cite{SinghAK2016}.
%Jpeg compression 

 Moreover, in the case of JPEG compression attack, it can be seen from Table \ref{tab:jpegcompressionattackcomparaison2} that our method gives better results than the approach in \cite{RefJ15}. In addition, as depicted in Table \ref{tab:Robustnesscomparisonforlena} and Table \ref{tab:RobustnesscomparisonMTAPmethodsforlena}, it can be observed that the proposed approach outperforms schemes 
 %\cite{lien2006watermarking},
  \cite{RefJ14}, \cite{RefJ15} and \cite{RefJ16}. In addition, the proposed method shows relatively good robustness to JPEG compression when the quality factor is $50$ except for the schemes \cite{RDWTMTAP2017}, \cite{SinghAK2015} and \cite{SinghAK2016} which outperform the proposed method is this particular case.  
%Cropping
\\ For cropping attack , according to Table \ref{tab:Robustnesscomparisonforlena}, the proposed scheme, compared with the other schemes, gives the best performance.
%Rotation attack
Moreover, it can be seen from Table \ref{tab:Robustnesscomparisonforlena} that the results obtained after rotation attack are encouraging. 
In our method, the NC values are not so good for the rotation attacks with degree greater than $\pm 0.75$; but it is far better than those in the listed methods (see Table \ref{tab:Robustnesscomparisonforlena}).
 \\ Furthermore, it can be concluded that the proposed method shows good robustness against several kind of attacks compared with alternative methods.

 \begin{table}[H]
\centering
\caption{NC values after Salt \& pepper noise attack}
\label{tab:SPNComparison}
\begin{tabular}{ccc}
\hline \noalign{\smallskip}
\: Salt \& pepper \: & \:  Proposed Scheme \: & \: Scheme \cite{RefJ15} \: \\
 \hline
$(\mu=0, \sigma=0.01)$ & $1.0$ & $0.83$%k=5000
\tabularnewline%k=16000(valeur optimale)

$(\mu=0, \sigma=0.02)$ & $0.9843$ & $0.76$ \tabularnewline% valeur optimale est k=6000
\noalign{\smallskip} \hline
\end{tabular}
\end{table}

\begin{table}[H]
\footnotesize
\centering
%{\renewcommand{\arraystretch}{1.2} 
\caption{NC values after JPEG compression with several quality factors }
\label{tab:jpegcompressionattackcomparaison2}
%R??sultats pour l'image baboon
\begin{tabular}{ccc}
\hline \noalign{\smallskip}
\: JPEG Compression \: & \:  Proposed Scheme  \: & \: Scheme \cite{RefJ15} \: \\
 \hline
 (QF=100) & 1.0 &0.96\\
(QF=90) & 1.0 & 0.95  \\
(QF=80) & 1.0 & 0.87 \\
(QF=70) & 0.9980 & 0.86\\
(QF=50) & 0.9384 & 0.79 \\
(QF=20) & 0.7382 & 0.66 \\
(QF=10) & 0.6545 & 0.61 \\
\noalign{\smallskip}\hline
\end{tabular}
\end{table}

\begin{table}[H]
%\footnotesize
\centering
%{\renewcommand{\arraystretch}{1.6}
%{\setlength{\tabcolsep}{0.4cm} 
\caption{Comparison of the robustness of the proposed algorithm with different methods against several attacks for Lena}
\label{tab:Robustnesscomparisonforlena}
\begin{tabular}{ccccc}
\hline
 & \multicolumn{4}{c}  {Watermarking methods}     \tabularnewline
\cline{2-5}
  Attacks & \: Our method \:  & \cite{RefJ14} &\cite{RefJ15}& \: \cite{RefJ16} \: \\ 
\hline \noalign{\smallskip}
%No Attack  & Infinity & \textbf{1.0} &&&&\tabularnewline  
%\hline
  HE & 1.0 & NA & 0.83& 0.79	 \\
\hline
%  Gaussian noise ($\mu=0 $  $ ,\sigma=0.001$) & 1.0  &  &&&&\tabularnewline
%\hline
 JPEG  & &  &&\\
  QF=70 & 1.0  & 0.51 & 0.86&  1.0   \\
  QF=90 & 1.0  & 1.0 & 0.95& 1.0   \\
  \hline
   Cropping &   &&&\\ 
  ($25\%$) & 1.0  & -- & 0.96& 0.60 \\  
  ($50\%$) & 0.9742  & -- & 0.90& -- \\  
  \hline
  Rotation  && &&\\
  $\theta = 0.25\degree$ & 1.0   & 0.37 & 0.75& 0.61\\
  $\theta = 0.75\degree$ & 0.9999  &0.26&0.67& 0.34\\
  $\theta = 1.0\degree$ & 0.55  &0.24&--& 0.27\\
  $\theta = -0.25\degree$ & 1.0  &0.32&0.76& 0.65 \\
  $\theta = -0.75\degree$ & 0.9998 &0.10&0.24&0.67\\
  $\theta = -1.0\degree$ & 0.57  &0.16&--& 0.28\\
  \noalign{\smallskip} \hline
\end{tabular}
\end{table}
\begin{table}[H]
\caption{Comparison of the robustness of the proposed algorithm with  Singh's method \cite{RDWTMTAP2017} against several attacks for Lena}
\label{tab:RobustnesscomparisonWithMTAP2017forlena}
%
%\footnotesize
\centering
%{\renewcommand{\arraystretch}{1.6}
%{\setlength{\tabcolsep}{0.4cm} 
\begin{tabular}{ccc}
\hline
 & \multicolumn{2}{c}  {Watermarking methods}     \tabularnewline
\cline{2-3}
  Attacks & \: Our method \:  & \cite{RDWTMTAP2017} \\ 
\hline \noalign{\smallskip}
%No Attack  & Infinity & \textbf{1.0} &&&&\tabularnewline  
%\hline
  Histogram equalization & 1.0 &  0.9902	 \\
\hline
%  Gaussian noise ($\mu=0 $  $ ,\sigma=0.001$) & 1.0  &  &&&&\tabularnewline
%\hline
 JPEG  & &\\
  QF=50 & 0.9587  & 0.9951   \\
    \hline
   Gaussian noise &   &\\ 
  ($\mu=0, \sigma=0.001$) & 0.9972  & 0.9965 \\  
  
  \hline
  Salt \& pepper noise & & \\
($\mu=0, \sigma=0.001$)& 0.9932& 0.9912\\
 \noalign{\smallskip} \hline
\end{tabular}
\end{table}

\begin{table}[H]
%\footnotesize
\centering
%{\renewcommand{\arraystretch}{1.6}
%{\setlength{\tabcolsep}{0.4cm} 
\caption{Comparison of the robustness of the proposed algorithm with different methods against several attacks for Lena}
\label{tab:RobustnesscomparisonMTAPmethodsforlena}
\begin{tabular}{ccccc}
\hline
 & \multicolumn{4}{c}  {Watermarking methods}     \tabularnewline
\cline{2-5}
  Attacks & \: Our method \:  & \cite{RDWTMTAP2017}&  \cite{SinghAK2015}  & \cite{SinghAK2016} \\ 
\hline \noalign{\smallskip}
%No Attack  & Infinity & \textbf{1.0} &&&&\tabularnewline  
%\hline
  Histogram equalization & 1.0 &  0.9902	&0.9208&0.9942 \\
\hline
%  Gaussian noise ($\mu=0 $  $ ,\sigma=0.001$) & 1.0  &  &&&&\tabularnewline
%\hline
 JPEG  & &&&\\
  QF=50 & 0.9384  & 0.9951 & 0.9994& 0.9935 \\
    \hline
   Gaussian noise &   &&&\\ 
  ($\mu=0, \sigma=0.01$) & 1.0 & 0.9965&0.9754&0.9828 \\  
  ($\mu=0, \sigma=0.5$) &  0.9803 &0.9865 &0.6565&0.8481 \\ 
 
  \hline
  Salt \& pepper noise & & &&\\
($\mu=0, \sigma=0.001$)& 0.9997& 0.9912&0.9952&0.9867\\
 \noalign{\smallskip} \hline
\end{tabular}
\end{table} 
\subsubsection{Complexity analysis comparison}
To further evaluate the computational complexity of the proposed method, a comparison with various related works \cite{RDWTMTAP2017}\cite{SinghAK2015}\cite{SinghAK2016} has been elaborated. The computational complexity of each method is expressed  in big $O$ notation. The scheme  in \cite{RDWTMTAP2017} is based on NSCT-RDWT-SVD and Arnold transform while scheme in \cite{SinghAK2015} used DWT, DCT and SVD transforms.  The proposed schemes in \cite{RDWTMTAP2017}\cite{SinghAK2015} have cubic complexity because of the use of SVD. The complexity of the scheme in \cite{SinghAK2016} is $O(n^2)$ due to Arnold transform. Our cubic complexity is due to the matrix multiplication performed during image reconstruction in the embedding process (equation \ref{eq:Imagereconstruction}). Although the complexity of the proposed method is much higher than scheme in  \cite{SinghAK2016}, both the accuracy of watermark extraction and the robustness of the proposed scheme are much more important than scheme in \cite{SinghAK2016}. 
%\begin{table}[H]
%\footnotesize
%\centering
%%{\renewcommand{\arraystretch}{1.2}
%\caption{Computational complexity comparison between the proposed scheme and schemes in \cite{RDWTMTAP2017}\cite{SinghAK2015}\cite{SinghAK2016}}
%\label{tab:ComplexityComparison}
%%R??sultats pour k=750
%\begin{tabular}{cc}
%\hline \noalign{\smallskip}
% \: Watermarking scheme \: & \: Complexity \\
% \hline
%% \cite{lien2006watermarking} &  41.54  \tabularnewline
%\cite{RDWTMTAP2017} & $O(N^3 )$  \\
%\cite{SinghAK2015} & $O(N^3)$\\
%\cite{SinghAK2016} &  $O(N^2)$\\
%Proposed method &  $O(N^3)$\\
%\noalign{\smallskip} \hline
%\end{tabular}
%\end{table} 
%\textcolor{blue}{
%It can be concluded from table \ref{tab:ComplexityComparison} that the complexity of  the proposed scheme is cubic.
% We note that the main computational component in our proposed technique is the the matrix multiplication in equation \ref{eq:Imagereconstruction} used to reconstruct the watermarked image. Compared with \cite{RDWTMTAP2017} and \cite{SinghAK2016} it  the complexity of our method is much higher. Nevertheless, the accuracy of the watermark extraction as well as the robustness of the proposed method against different attacks is much more important than schemes \cite{RDWTMTAP2017} and \cite{SinghAK2016}.  } 
\section{Conclusion}
\label{Conclusion}
In this work, a blind robust hybrid image watermarking scheme combining the two well known transformations DFT and DCT for Copyright protection is presented. The watermark is embedded in the middle band DCT coefficients of the DFT magnitude of the cover image using two secret keys for increasing security. The first one is used to generate the PN sequences to be inserted in the watermark embedding while the second one is to encrypt the watermark with Arnold transform. Taking the advantages of jointing DFT and DCT transforms, the  obtained results show that the proposed scheme ensures good resistance to a wide variety of attacks for textured images as well as natural images while preserving high imperceptibility. Future work can be focused on investigating the proposed technique for another kind of image and enhancing its robustness against new variety of attacks.
\bibliographystyle{spbasic}      % basic style, author-year citations


\end{document}